\documentclass[usenatbib]{mn2e}
\usepackage{longtable}
\usepackage{rotating}
\usepackage{multicol}
\usepackage{multirow}
\usepackage{rotating}
\usepackage{amssymb}
\usepackage{amsmath}
\usepackage{times}
\input psfig
\def\ltsim{\raise 2pt \hbox {$<$} \kern-1.1em \lower 4pt \hbox {$\sim$}}
\def\ltapprox{\raise 2pt \hbox {$<$} \kern-1.1em \lower 5pt \hbox {$\approx$}}
\def\gtsim{\raise 2pt \hbox {$>$} \kern-1.1em \lower 4pt \hbox {$\sim$}}
\def\gtapprox{\raise 2pt \hbox {$>$} \kern-1.1em \lower 5pt \hbox {$\approx$}}

\title[Radio Survey of Southern X-ray Luminous Clusters]{A Radio Survey 
of Seven Southern X-ray Luminous Clusters of Galaxies}
\author[O. B. Slee et al.]
{O.B.~Slee,$^{1}$\thanks{email: Bruce.Slee@csiro.au} H.~Andernach,$^2$ 
V.J. McIntyre,$^{1,3}$ G.~Tsarevsky\,$^{1,4}$ \\
$^1$Australia Telescope National Facility, CSIRO, P.O. Box 76, Epping,
NSW 1710, Australia\\
$^2$Departamento de Astronom\'{i}a, Universidad de Guanajuato, AP 144,
Guanajuato CP 36000, Mexico\\
$^3$School of Physics, University of Sydney, NSW 2006, Australia \\
$^4$Astro Space Center, Lebedev Physical Institute, RAS, Moscow, 117910 Russia \\
}
\date{MNRAS Submitted, }

\begin{document}



\maketitle


\begin{abstract}
The Australia Telescope Compact Array (ATCA) has been used at 1.38 and 
2.38~GHz to survey  seven southern Abell clusters of galaxies with high X-ray
luminosities: A2746, A2837, A3126, A3216, A3230, A3827 and A3836. The
clusters have also been surveyed at 0.843~GHz with the Molonglo Observatory
Synthesis Telescope (MOST). We have listed a complete 1.38-GHz sample of 149 radio
sources within the Abell circles centred on their X-ray centroids. We
compare their identification fractions, emitted 1.38-GHz and optical
powers, radio spectral indices and radial variation in projected source
density with those of the radio-selected samples of Slee et al.\ (1998). We
compare our fractional radio luminosity function with that of the radio-selected
samples of Ledlow \& Owen (1996) and Slee et al.\ (1998). Three significant
differences are noted between X-ray and radio-selected samples of
clusters; (1) the X-ray sample has an excess of flat-spectrum radio
sources; (2) the fractional radio luminosity function for the FR\,I sources
in the X-ray selected sample is much steeper, implying that  fewer of
their cluster galaxies  become hosts for the stronger FR\,I radio galaxies;
(3) a complete absence of FR\,{\sc II} radio galaxies in the X-ray selected sample.
The average excess projected density of radio sources near our cluster centres 
is $\sim$5 times the background source density.
\end{abstract}


\section{Introduction}   \label{intro}

Clusters of galaxies provide an interesting environment in which to
test theories of how radio galaxies form and evolve. In particular, the
presence of a relatively dense, X-ray emitting ionised gas permeating
clusters provides an ideal laboratory to study jet propagation and the 
expansion of radio lobes.

The examination of these processes requires carefully-defined  samples
incorporating data on the individual cluster members from the radio,
optical and X-ray regimes. Further, selection of the cluster samples may
have an impact upon the conclusions that can be drawn. Clusters selected
primarily on the radio power of the objects within them may represent
a distinct environment, with a radio source population different from that in
clusters selected on the basis of integrated optical luminosities or
X-ray luminosity.

A number of synthesis radio surveys of large samples of clusters have
been published in the last 17 years. In addition, numerous publications
on individual clusters and individual cluster radio galaxies exist
in the literature. Comprehensive collections of 1.4-GHz maps of
cluster radio galaxies taken with the Very Large Array (VLA) of the 
National Radio Astronomy Observatory (NRAO, USA) were published by 
Zhao, Burns \& Owen (1989),
Owen et al.\ (1992, 1993), Ledlow \& Owen (1995a), and a detailed analysis
of these data were made by Ledlow \& Owen (1995b, 1996). A second set of
VLA maps at 1.5 and 4.9~GHz with complete source lists were published
by Slee, Perley \& Siegman (1989), Slee, Roy \& Savage (1994) and Slee,
Roy \& Andernach (1996), who later drew general conclusions about the
relationships  between radio, optical and X-ray parameters (Slee, Roy \&
Andernach 1998). A comprehensive survey at 0.843~GHz of 39 nearby southern 
clusters was made by Unewisse (1993), but none of the more distant 
clusters in the present paper was included.

The present paper presents complete source lists and analysis for seven
rich clusters from Abell et al's (1989) catalogue south of Declination 
$-50^{\circ}$, selected on account of their high ROSAT X-ray luminosities 
in the ``X-ray brightest Abell-type clusters of galaxies'' (XBACs) 
catalogue of Ebeling et al.\ (1996). The radio survey was made
at 1.38 and 2.38~GHz with the Australia Telescope Compact Array (ATCA)
with the initial object of identifying a suitable candidate cluster
in a search for radio gravitational lensing (Tsarevsky et al.\ 2008, in
preparation). In addition, as an X-ray-selected sample, the source list
and its general analysis are of interest in their own right.

Our selection criteria included the following:

\noindent
1.~Clusters south of Declination $-50^{\circ}$, enabling observations in
the ``cut'' mode (cf.\ Sect.\ 2.1) with good coverage of the uv plane.

\noindent
2.~Cluster X-ray luminosities (0.1--2.4~keV) above the median of 
1.56$\times10^{37}$\,W of the XBACs sample; the exception is A3126, 
whose $L_X$ falls only 13\% below the median, and was included because 
it has a well-determined redshift and a high velocity dispersion.

\noindent
3.~Contamination of the ROSAT image by active galactic nuclei (AGN) is minimal 
as assessed by Ebeling et al.\ (1996).

The clusters were also surveyed with the Molonglo Observatory Synthesis Telescope
(MOST) at 0.843~GHz with the primary object of detecting as many as possible
of the 1.38 and 2.38-GHz sources in order to obtain more accurate radio
spectral indices and, if possible, identify steep-spectrum radio relics,
which are known to favour the cluster environment.

Section~2 briefly describes the observations, while Section~3 presents
the radio source lists and optical identifications. Section~4 discusses
the derived parameters of the sources, including spectral indices,
relationships between emitted radio, optical and X-ray luminosities and
the clusters' radio luminosity function, and Section~5 presents a
discussion of our results.

In the following text, frequent reference is made to cluster samples
selected by X-ray emission, radio emission or using neither of these
criteria. To avoid confusion, we call samples selected by their X-ray
emission ``X-ray samples'' and those selected by their radio emission
are termed ``radio samples''; naturally, both samples imply that optical
clusters are present.

For consistency with previously published radio and X-ray data, to which 
we refer in this paper, we use the Einstein-de Sitter cosmological model with
H$_0$=75\,km\,s$^{-1}$\,Mpc$^{-1}$ and q$_0$=0.5 throughout this paper.

\section{Observations}  \label{obs}    

\subsection{The ATCA observations}  \label{ATCAobs}  

The general survey of the seven clusters involved simultaneous 1.38
and 2.38-GHz observations with the ATCA in the 1.5C and 1.5D configurations 
over two 15-hour sessions in February and May 1996. 
We utilised the standard continuum correlator configuration which
has dual frequency mode with two independent 128 MHz bands.
One of the target clusters, A3827, was subsequently selected for much
deeper observations at 4.80 and 8.64~GHz in order to search for evidence of
gravitational lensing; this is treated separately in Tsarevsky et al.\
(2008). Table~1 lists the more useful optical and X-ray data for these
clusters, and the rms radio noise levels over the cleaned maps.

\begin{table*}
\caption{The list of surveyed clusters.
The 843-MHz noise levels pertain to the image centre.  The higher rms for 
A3216 is caused by residual sidelobe structure of a 25-Jy source lying 
0.7$^{\circ}$ from the X-ray centre (SUMSS~J040820$-$654508).}  
\begin{tabular}{lccclrlcccccc}
\hline
Cluster & R  &  BM  & RA,DEC(J2000,X-ray)  & $z_{LG}$ & $N_z$ & Ref. & $\sigma_V$ & $R_A$ & L$_X$ & \multicolumn{3}{c}{$\sigma_{radio}$, mJy/beam} \\
        &    &     &  h~~m~~s~~~~~~$^{\circ}~'~~''$~~ref &  & & for~$z$ &  km\,s$^{-1}$  & $'$ & 10$^{37}$\,W  & 0.843 & 1.38 & 2.38\,GHz \\
\hline
A2746 & 0 & II-III & 00\,14\,18.4~$-$66\,04\,39 R & .1594 &  5 & 1      &      & 14.0 & 2.31 & 1.3 & 0.121 & 0.121 \\
A2837 & 0 &  I-II  & 00\,52\,44.9~$-$80\,15\,59 R & .1134 &  7 & 1,2    &      & 18.3 & 2.77 & 1.4 & 0.132 & 0.132 \\
A3126 & 1 &  III   & 03\,28\,37.5~$-$55\,42\,46 R & .0854 & 44 & 3,4    & 1020 & 23.2 & 1.36 & 1.1 & 0.085 & 0.048 \\
A3216 & 2 & II-III & 04\,04\,07.0~$-$65\,12\,32 X & .1581 &  1 & 5      &      & 14.0 & 2.23 & 2.8 & 0.080 & 0.093 \\
A3230 & 2 &  II    & 04\,11\,20.4~$-$63\,41\,46 X & .162~~e&-- & 6      &      & 13.8 & 2.43 & 2.4 & 0.072 & 0.048 \\
A3827 & 2 &  I     & 22\,01\,56.0~$-$59\,56\,58 R & .0983 & 21 & 7,8,10 & 1093 & 20.6 & 3.61 & 1.1 & 0.129 & 0.129 \\
A3836 & 2 &  I     & 22\,09\,23.3~$-$51\,48\,54 R & .1100 & 12 & 9      &  510 & 18.7 & 1.72 & 1.4 & 0.126 & 0.126 \\
\hline
\end{tabular}

\begin{flushleft} The cluster name is followed by the
Abell richness class R, the Bautz-Morgan (BM) type, the X-ray position taken from
B\"ohringer et al.\ (2004, REFLEX, marked ``R") when available, otherwise from
Ebeling et al.\ (1996, XBACs, marked ``X''). The galactocentric mean redshift of the
cluster, the number of galaxies with measured $z$, references contributing to
the redshifts, the radial velocity dispersion, Abell radius, X-ray luminosity 
(average of values published in REFLEX and XBACs, converted to
H$_0$=75\,km\,s$^{-1}$\,Mpc$^{-1}$ and q$_0$=0.5. The last three columns
are the 1-$\sigma$ noise levels at the map centres achieved at 0.843, 1.38, and 2.38\,GHz.
Note that the higher noise level for the 0.843-GHz image for A3216 is due to a 
strong source south of the cluster.
Redshift data were merged from the following
references: (1) B\"ohringer et al.\ 2004;
(2) De\,Grandi et al.\ 1999;
(3) Colless \& Hewett 1987;
(4) Lucey et al.\ 1983;
(5) Jones et al.\ 2004 (6dF-DR1);
(6) estimate from Peacock \& West 1992 (priv.\ comm.\ M.~West);
(7) West \& Frandsen 1981;
(8) Katgert et al.\ 1998;
(9) Ebeling, H., 1997, priv.\ comm.;
(10) Jones et al.\ 2005 (6dF-DR2).
\end{flushleft}
\end{table*}

The survey, centred near the X-ray centroid in each cluster, was made
by taking 30-min integrations, which were preceded and followed by a
5-min observation of a nearby phase calibrator. Seven to nine 30-min
observations were made on each field during the time (typically 14\,h)
for which the cluster was above the telescope elevation limits. The
primary flux calibrator B1934$-$638 was observed at the beginning of each
session. The mapping, cleaning  and restoration were performed in the
MIRIAD package (cf.\ Sault et al.\ (1995) and www.atnf.csiro.au/computing/software/miriad). 
Self-calibration was utilized in three fields to reduce
the contamination by strong field sources. Care was taken at each
iteration of the CLEAN algorithm, both in the general cleaning and
self-cal process, to use only the number of iterations necessary to
achieve a stable level of cleaned flux, while ensuring that the rms
residual did not fall below the theoretical value due to the expected
system noise.  The rms noise residuals achieved near the centres of the
cleaned maps are given in Table~1, which shows that they varied between
0.121 and 0.132~mJy\,beam$^{-1}$, except for the self-calibrated maps
(A3126, A3216, A3230), in which the rms was significantly lower. The
angular resolution (FWHM) was typically $\sim18''$ at 1.38~GHz and
$\sim10''$ at 2.38~GHz.

\subsection{The MOST observations}   \label{MOSTobs} 

The MOST data were obtained from the archive at the University of Sydney's
Institute of Astronomy 
and from new observations carried out as part of the Sydney University 
Molonglo Sky Survey (SUMSS; see Bock, Large \& Sadler, 1999 and Mauch et 
al.\ 2003). The SUMSS data are available, and fully described, at the 
project web site, http://www.physics.usyd.edu.au/ioa/Main/SUMSS.
The archival data were obtained when one of us (V.\ McIntyre) was a
research fellow in the School of Physics. Those observations were 
carried out by R.W.\ Hunstead and J.G.\ Robertson in the 1990s as part of 
a long-term program of cluster observations known as the Molonglo Cluster 
Survey (Haigh et al.\ 1997, Haigh 2000).  The archival observations were 
taken in a mode which provides a 70$'$ diameter field, and the new 
observations in a wide-field mode, where the field of view has a diameter 
of 2.3$^{\circ}$. The angular resolution (FWHM) was $\sim45''$;
the rms noise level for each cluster field is listed in Table~1.

All the observations were reduced using the latest version of the
pipeline-processing software for the SUMSS. Briefly, this involves
determining calibration factors, back-projecting the intensities recorded
by the telescope beams (Perley 1979), and CLEANING to reduce the sidelobe 
structure. A more complete description of this software system is
presented by Cram \& Ye (1995).

\section{Results}  \label{results}  

\subsection{Measurement of source parameters} \label{errors}  

The rms residuals of the Gaussian fitted positions, i.e.\ ~$[(\Delta$\,RA)$^2
+(\Delta$\,Dec)$^2)]^{1/2}$ depend on several factors including the signal/noise
ratio and the angular extent of the source. We have fitted  146 sources
listed at 1.38 GHz in Table~2  with unconstrained elliptical Gaussians;
the rms residuals of the Gaussian fitted positions ranged from 0.1$''$ to 10$''$, 
with a median of 0.9$''$, but their distribution was 
highly skewed with an interquartile distance of 0.9$''$ due to the influence 
of signal/noise and angular size.  The accuracy of the SuperCOSMOS  positions 
is generally better than $\pm$0.4$''$ (Hambly et al.\ 2001), and so is 
not an important influence in the identification procedure. The 0.843-GHz 
positions, being derived from 3 to 5-times larger restoring beams were 
utilized only as an aid to match 0.843-GHz sources to corresponding 1.38 
and 2.38-GHz sources.

The adopted flux density for a source was computed on the basis of two 
of three possible methods. For an unresolved source, we averaged the peak
flux densities of point-source and unconstrained gaussian fits. For
a resolved source, we averaged the integrated flux densities from
an unconstrained gaussian fit and from adding pixels in a small box
enclosing the full source extent.

The adopted flux density for an unresolved source at 1.38 and 2.38~GHz was
generally a mean of the results from the three methods of measurement;
for a resolved source, we averaged the integrated fluxes from the
unconstrained Gaussian fit and the pixel-adding method.

The errors in flux density at 1.38 and 2.38~GHz due to system noise and
sidelobes were assessed from the rms residuals in the peak flux from
the two methods of Gaussian fitting (point source and unconstrained Gaussian), 
usually an average being taken for unresolved sources. If the source was
resolved, we adopted the error in the peak flux from the unconstrained
Gaussian fit. To this error, we added in quadrature a systematic error of
6\% of the derived flux density, as explained in detail by Slee et al.\
(1996).

We adopted a similar method of deriving 0.843-GHz flux densities and
their errors. Since there were very few resolved sources at 0.843~GHz,
we almost exclusively adopted the peak flux density and error from the
point-source Gaussian fit, but we ensured that the two other methods
gave consistent values. A systematic error of 3\% of the 0.843-GHz flux
density was added in quadrature with the noise and sidelobe error
(Mauch et al.\ 2003).

\subsection{The source list}  \label{sourcelist}  

Table~2 presents the parameters determined for the 149 radio sources
detected in the fields of the seven southern clusters. Including subcomponents
and integral parameters of double sources, Table~2 contains a total of
168~entries.  The 1.38-GHz flux densities are the primary measurements,
since this is the survey with highest sensitivity. The 0.843-GHz and
2.38-GHz flux densities are used to complement the 1.38-GHz intensities in
order to provide spectral indices for the majority of radio sources. The
2.38 and 1.38-GHz measurements also provide angular sizes for many of
the sources. Only two sources were detected exclusively at 2.38~GHz,
and no extra sources were detected at 0.843~GHz within the area 
defined by the 1.38 GHz observations.

The sources in Table~2 are drawn from areas on the 1.38-GHz maps that
are within the 32-arcmin primary beam to FWHM (3~clusters) or moderately
outside the FWHM circle (4~clusters).
We consider that the 149~sources (omitting components of doubles) in 
Table~2 constitute a complete sample to a 1.38~GHz flux density of 1.0~mJy. 
We justify this by noting that a 1.38-GHz source of 1\,mJy at one Abell 
radius from the pointing center varies between four and eight times the rms
noise, depending on the Abell radius of the cluster. At 2.38~GHz, a 
1-mJy source near one Abell radius and beyond is usually undetectable 
unless it is reasonably strong and/or has a flat spectrum, but 
we have quoted an upper flux limit that is five times a 
measured rms in the region of the source. If no 2.38-GHz flux density 
or upper limit is listed in column~9, the source is outside the cleaned inner
quarter of the dirty map.

The spectral indices, $\alpha$, in Table~2, defined by S($\nu$) $\propto
\nu^{\alpha}$ were derived from the flux densities or their upper
limits at the three frequencies; for some of the stronger sources
additional data from higher frequency surveys were extracted from the
literature, using the CATS database (cats.sao.ru, see Verkhodanov et
al.\ 1997). Angular sizes were derived by fitting elliptical Gaussians
to the restored images, and were accepted only if the major
and minor axes significantly ($>2\sigma$) exceeded the major and minor
axes of the restoring beam; sometimes the angular sizes were available
at both 1.38 and 2.38~GHz, in which case the tabulated value is the
average. For double sources, indicated by ``a/b'' in the source number,
we tabulated the angular separation between the fitted positions of the
components and the position angle of the vector (from north through east)
from the component~(a) of higher flux density.

As an example of the images obtained, Figures~1 and 2 show the 1.384 and
0.843~GHz contour maps of the region around the centre of Abell~3836
with the sources numbered according to their designations in Tables~2
and 3. The circle shown on the map has the Abell radius of the cluster,
which is 2.0~Mpc for the adopted cosmology. Similar maps were constructed
for the remaining six clusters, and the 1.38-GHz maps of these are 
available in the online material of the present paper.

\begin{figure*}
\psfig{file=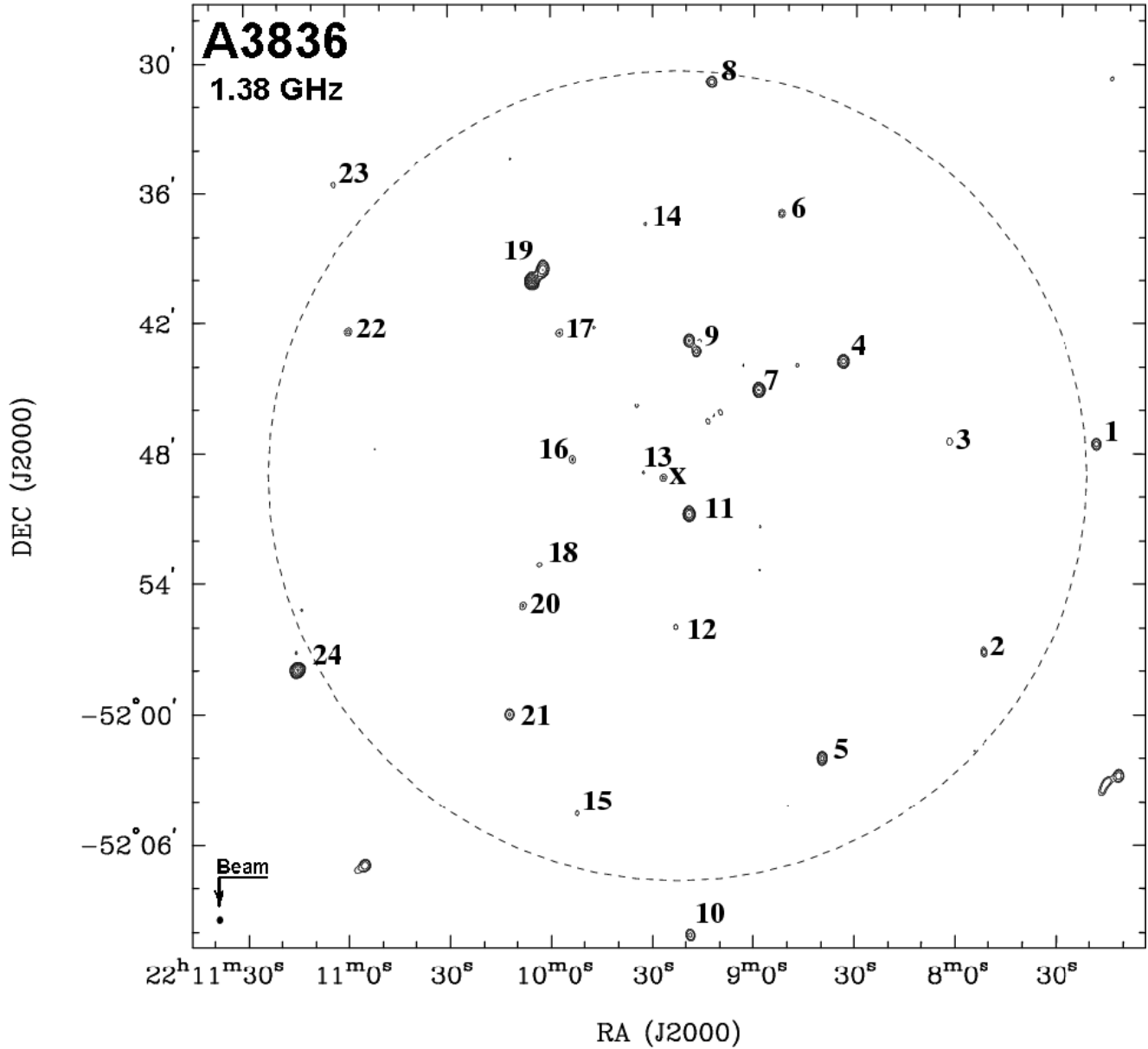,width=17.5cm,angle=0}
\caption{ \noindent  A 1.38-GHz cleaned map of A3836, obtained with ATCA 
(uncorrected for primary beam attenuation) centred on the X-ray centroid 
marked ``X'', and covering an area of 43$'\times43'$. The dashed circle marks 
the Abell radius (2.0\,Mpc for the adopted cosmology), and the radio sources are 
numbered according to their entries in Tables~2 and 3. The contour levels 
are -0.57, 0.57, 0.96, 1.9, 3.8, 7.7, 13.4, and 17.2 mJy\,beam$^{-1}$. The restoring beam, shown
in the lower left corner, has FWHM of 20.3$''\times16.9''$. The rms noise over
clear areas near the centre is  0.126~mJy\,beam$^{-1}$. The primary beam width
(FWHM) at 1.38~GHz is 32$'$.
}
\end{figure*}

\begin{figure*}
\psfig{file=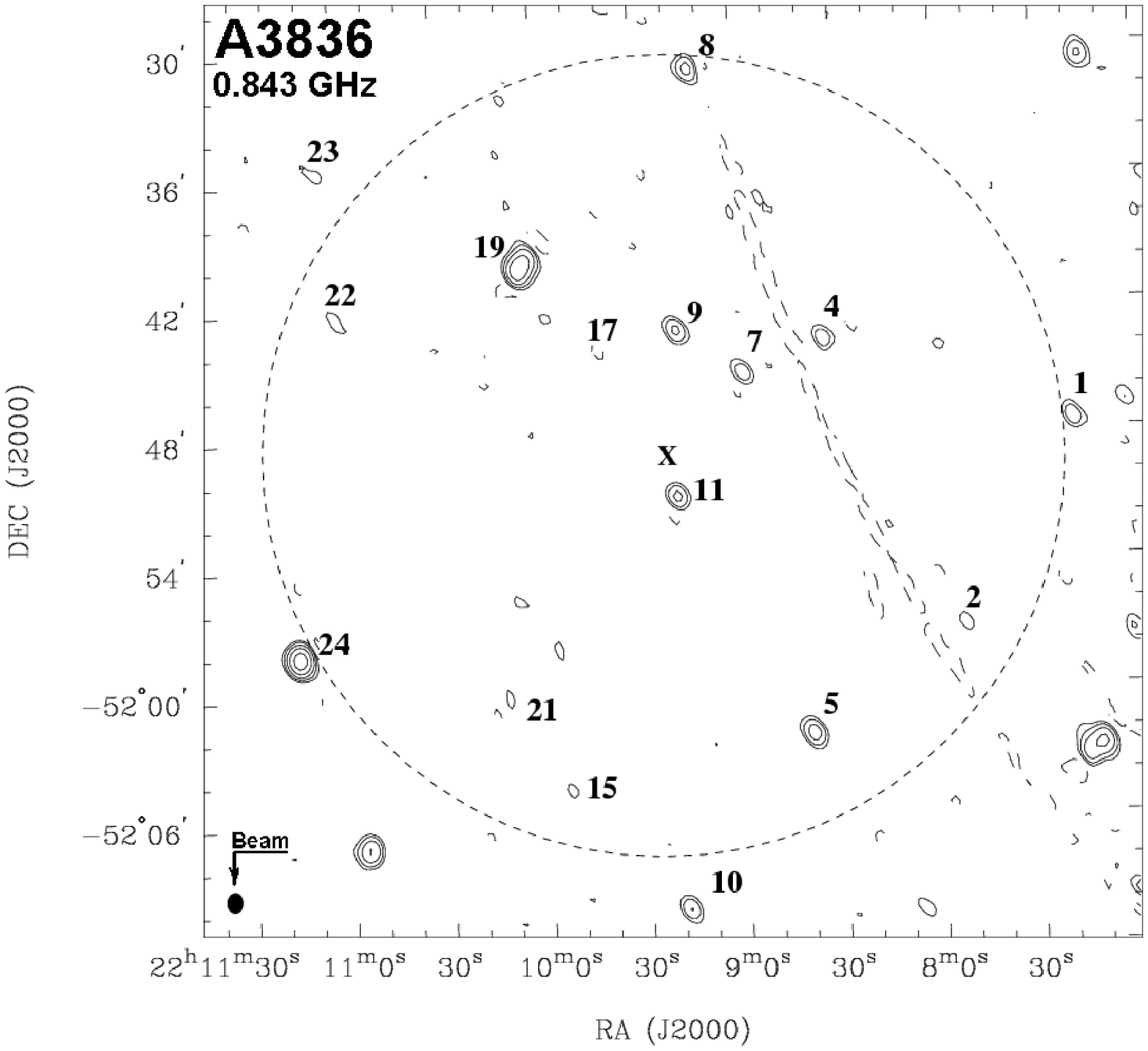,width=17.5cm,angle=0}
\caption{ \noindent  An 0.843-GHz map of A3836 obained with MOST, centred 
on the X-ray centroid marked X, covering the same area as that of Fig.~1.
The dashed circle marks the Abell radius (2.0\,Mpc),
and the radio sources are numbered according to their entries in Tables~2
and 3. The restoring beam, shown in the lower left corner, has FWHM of $\sim45''$. 
The rms noise over clear areas is 1.4~mJy~beam$^{-1}$. 
The contour levels are $-$3, 3, 6, 12, 24, 48 mJy~beam$^{-1}$.
The arc-like structure is a residual sidelobe of a source far outside
the field of view.  } 
\end{figure*}

\subsection{Optical identifications}  \label{optIDs}  

Each of the 149 sources in Table~2 was checked for an optical counterpart
in the SuperCOSMOS Science Archive (http://surveys.roe.ac.uk/ssa), based
on digitised data from the B, R and IR plates exposed at the UK Schmidt
Telescope, as well as data from the ESO~R plate. SuperCOSMOS offers an
accurate position for the object on each plate and estimates its apparent
magnitude corrected for atmospheric extinction. Interstellar extinction
was obtained from Schlegel et al.\ (1998). The redshifts attributed to
optical identifications in Table~3 were those available in published form
as of June~1, 2007.  We have included in Table~3 only those radio sources
for which we found an optical identification. We also include ten blank
fields (B) in Table~3, and give notes for all of them in this section,
and present overlays for nine of them in Figure~3.

We examined in detail the observational factors that were most
likely to influence our acceptance of an optical object as a likely
identification. These factors included the accuracies of the optical
and radio coordinates, the angular sizes of the optical objects and radio
sources, and the probability of a chance optical association in a
given circular area about the radio position. The accuracies
of the radio and optical positions have been discussed in Section\ 3.1.

SuperCOSMOS also classifies the objects on each plate as either a galaxy
(G) or star-like (St), according to their morphology. We accepted a
classification of either galaxy or star-like depending upon an assessment
of the four separate classifications backed up by an examination of the
individual red and blue digitized images, resulting in the derivation 
of the object's colour index.  We were able to classify the morphological
type of ``galaxy'' for the brighter objects, but in the case of faint
star-like objects it was not possible to discriminate between actual
stars, quasars, or faint, misclassified galaxies. The results from this
radio/optical comparison are presented in Table~3.

The brightness of a galaxy identification is important in deciding
whether the galaxy is a cluster member, especially as few of the suggested
identifications have measured redshifts. It is now recognized that if
an E/S0 galaxy is to be the host of a radio galaxy, its absolute red
magnitude is almost invariably M$_R\le-$21.0 (Ledlow \& Owen 1995b, Slee
et al.\ 1998). Therefore, in order to check whether the galaxy was a likely 
cluster member, we transformed the apparent $R$-band magnitude to an absolute 
$R$-band magnitude (including K, general dimming and extinction corrections) 
by the application of the mean cluster redshift. Thus we were able to
distinguish between cluster galaxies and radio sources behind the cluster,
but the method does not eliminate a foreground  identification. However,
Slee et al.\ (1998) have shown that most of the identifications that do
not satisfy the above condition are background objects, because their
distribution in $m_R$ has a median value 2.5~mag fainter than that of the
identifications that do satisfy the condition. There are, in fact, two
foreground galaxies with measured redshifts in Table~3. The star-like
(St) identifications were not checked by this method, as these objects may
be quasars, AGN, faint misclassified galaxies or, more rarely, real stars.

In interpreting the entries in Table~3, the following notes may be useful: \\
1. All galaxy identifications with measured redshifts are given their 
correct values of log\,P$_{1.38}$ and the $R$-band absolute magnitude, $M_R$, 
regardless of whether the redshift places the radio source and its optical 
identification within the designated cluster. If they are not cluster 
members, the values of log\,P$_{1.38}$ and $M_R$ carry a trailing 
letter ``n''. \\
2. Galaxy identifications without measured redshifts are first
checked for membership of the designated cluster by using the
cluster's mean redshift to test whether the galaxy's $M_R$ is brighter
than $-$20.6 (see the next paragraph). If its $M_R$ fulfills that
criterion, both log\,P$_{1.38}$ and $M_R$ are listed in Table~3, using the
designated cluster's mean redshift. \\
3. We accept that all the St objects listed as identifications are stars or 
AGN, and so do not quote values for radio power and absolute red magnitude 
for them. 

The last two columns of Table~3 contain the emitted radio powers and 
absolute $R$-band magnitudes of those galaxies with $M_R\le-$20.6, and thus 
accepted as cluster members. Their UKST $R_F$-band magnitudes have been 
transformed to the Cousins sequence by applying the small corrections 
tabulated by Frei \& Gunn (1994). In what follows we refer to these corrected
magnitudes as $R$-band magnitudes.

\begin{figure*}
\centerline{\psfig{file=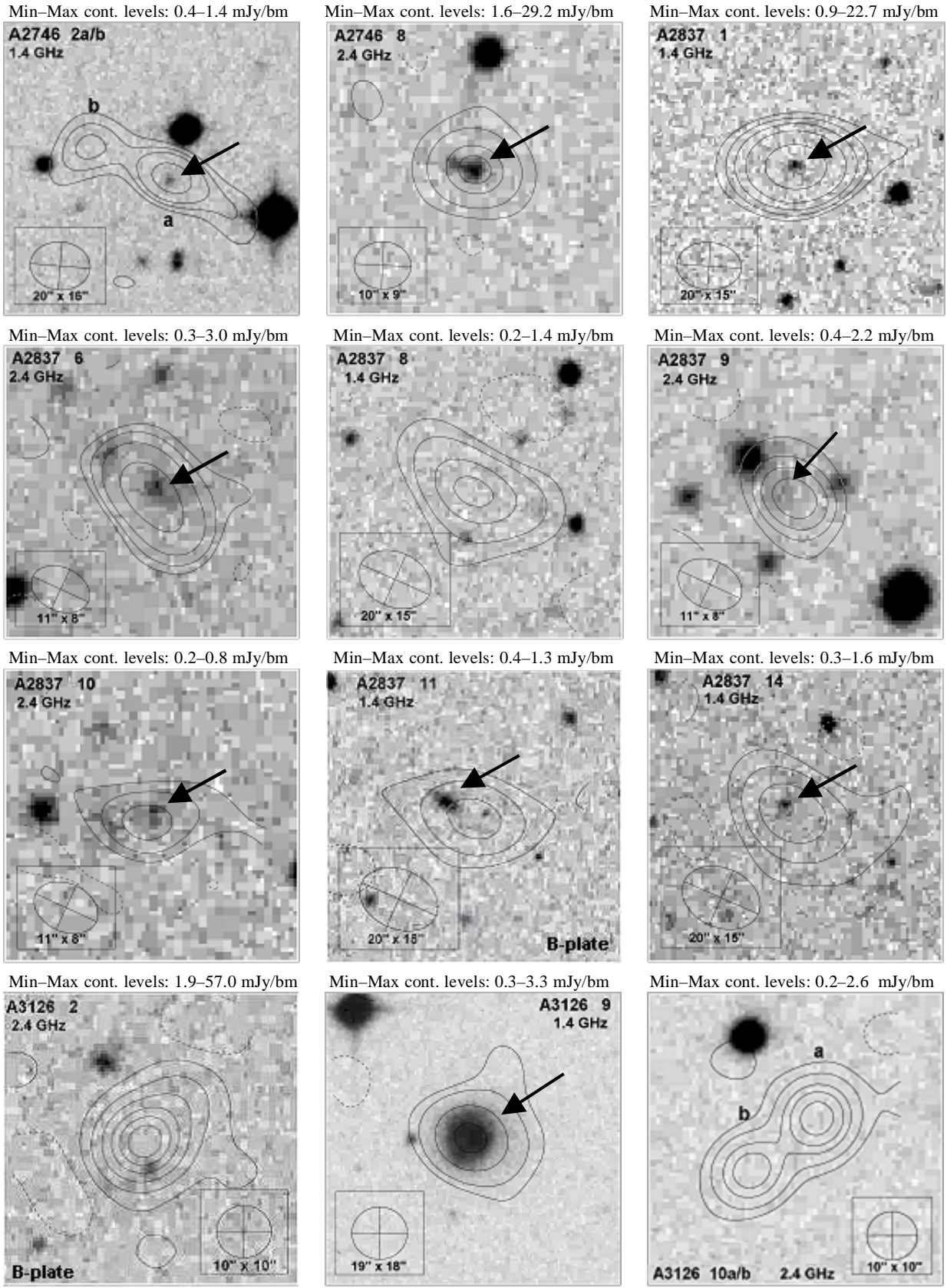,width=16.5cm,angle=0,bbllx=86pt,bblly=122pt,bburx=524pt,bbury=725pt,clip=}}
\caption{ \noindent Overlays of ATCA radio contours (corrected for primary
beam attenuation) on digitized images of the UKST second-epoch $R$-band sky survey
for sources with possible radio-optical identifications ($B_J$-plates were 
used for A3126\_2, A3827\_14 and A3836\_17). East is to the left and
North is to the top. The observing radio frequency and corresponding
restoring beam axes are indicated; minimum-maximum contour levels appear
above each image. See Tables~2 and 3 as well as the notes to individual
sources (Section~3.4).  } 
\end{figure*}

\setcounter{figure}{2}
\begin{figure*}
\centerline{\psfig{file=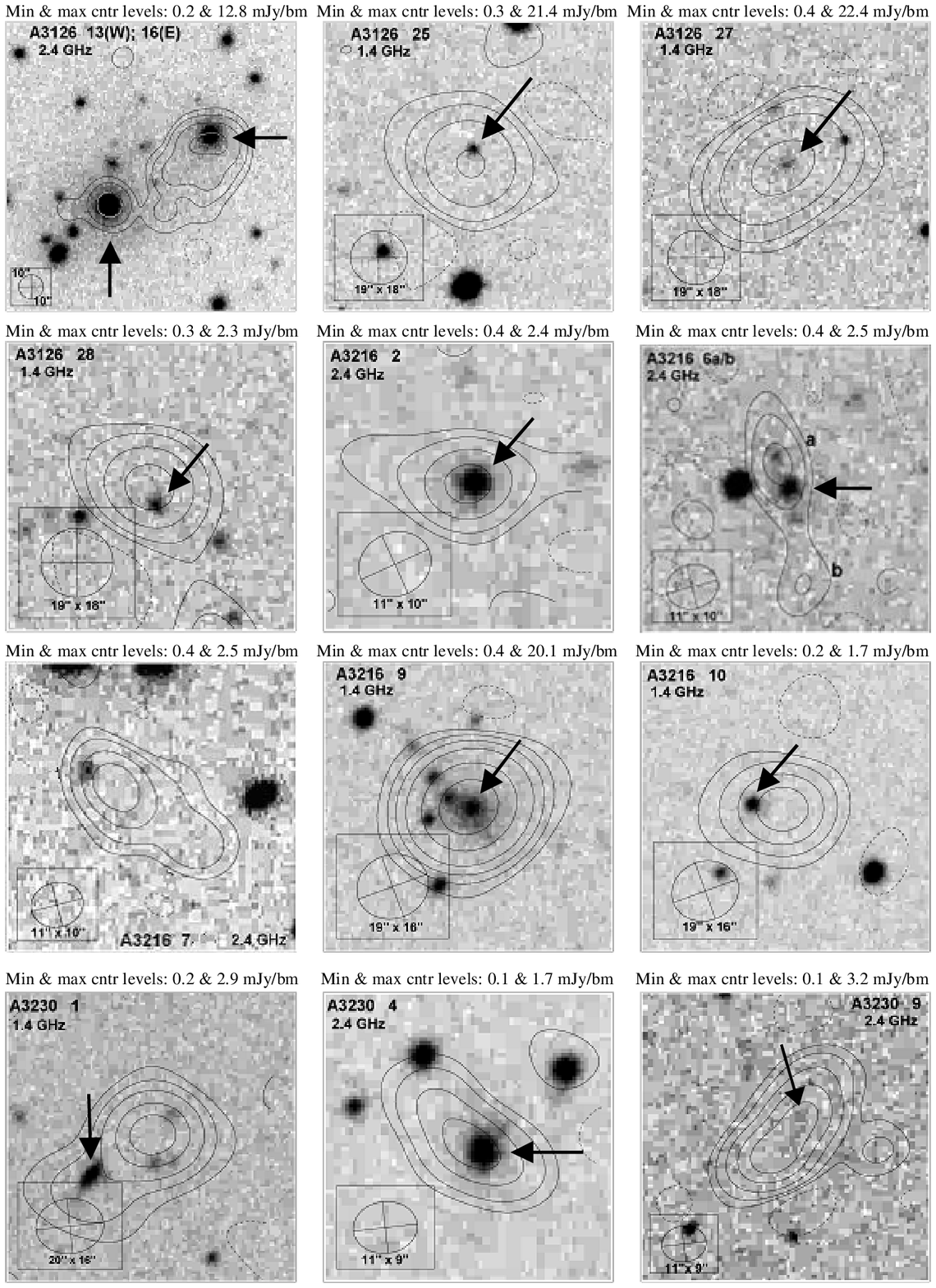,width=17.cm,angle=0,bbllx=84pt,bblly=116pt,bburx=524pt,bbury=725pt,clip=}}
\caption{ \noindent -- continued }
\end{figure*}

\setcounter{figure}{2}
\begin{figure*}
\centerline{\psfig{file=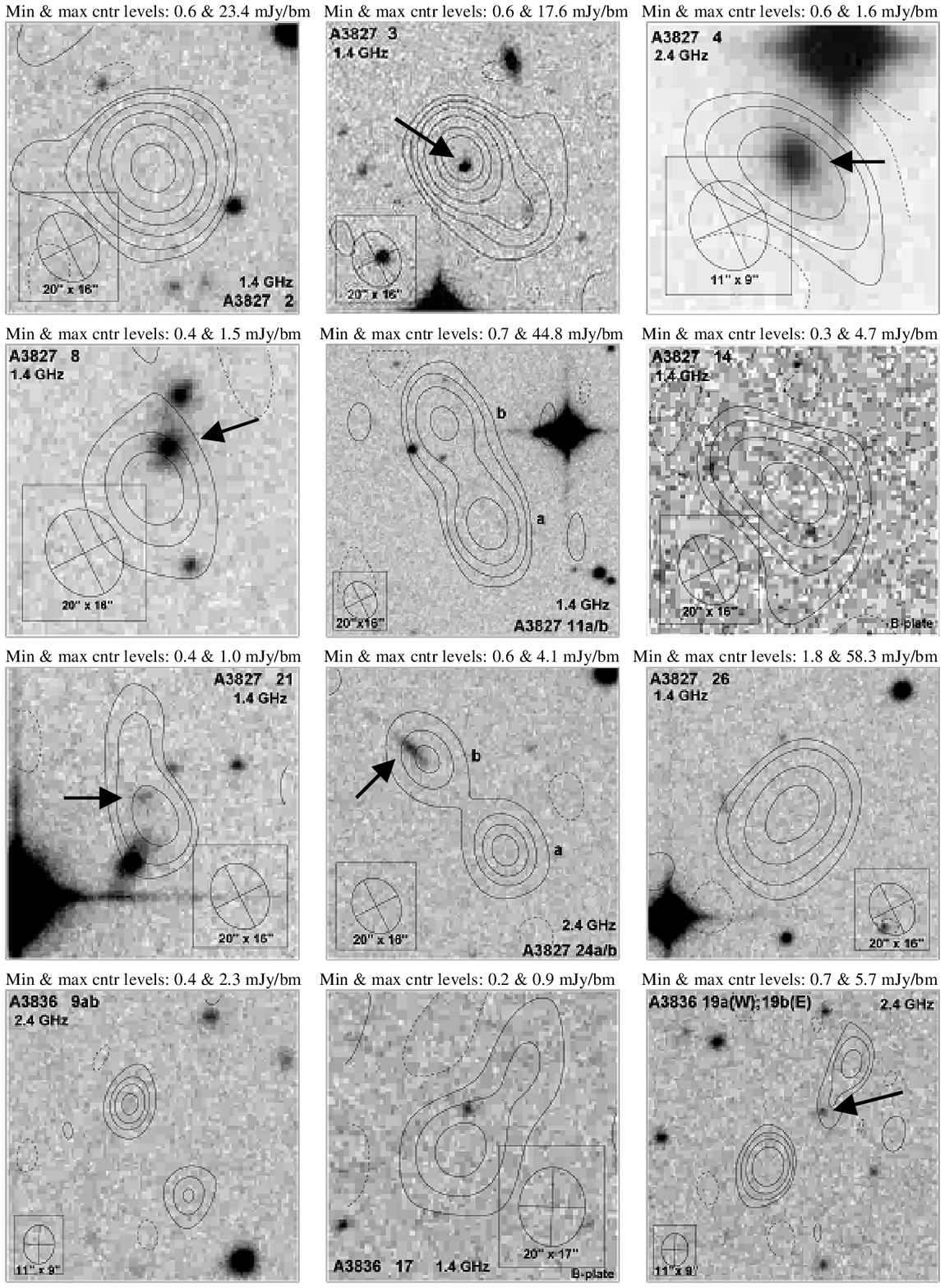,width=17.cm,angle=0,bbllx=84pt,bblly=122pt,bburx=526pt,bbury=724pt,clip=}}
\caption{ \noindent -- continued }
\end{figure*}

\subsection{Radio-optical overlays and notes to individual objects} \label{overlays}  

The montage of Figure~3 presents a series of radio-optical overlays, which
illustrate the application of our adopted identification procedure, and 
illustrates a few of the more interesting images of complex sources. 
If available with sufficient signal/noise, the higher resolution 2.38-GHz
contours, otherwise the 1.38-GHz contours, are overlayed on digitised plates 
from the UK Schmidt Telescope Red (IIIaF) Southern (Second Epoch) Survey
(see www-wfau.roe.ac.uk/sss/surveys.html).  Reference to Table~3 will help in 
interpreting these overlays.

First, in many cases we select a galaxy close to the radio centroid. These 
galaxies, e.g.\ in the frame for A2746\_2a/b (see note below), may be as
faint as $m_R$=19.65, and often there is no alternative optical object within the
radio contours. We do not consider such faint galaxy identifications as
members of the clusters on which they are projected because they violate
our absolute $R$-band magnitude criterion discussed in Section~\ref{optIDs}.

In some frames of the montage there are multiple objects within the radio
contours. Usually (e.g.\ A3216\_9), we select the brightest galaxy
near the radio centroid as the more likely identification.  In more ambiguous 
frames (e.g.\ A3230\_1) we accepted a brighter galaxy further away 
from the radio peak as the most likely identification, for reasons explained
in the individual notes below.

Several of the stronger radio sources, e.g., A3827\_6 and A3836\_9a/b (see below)
do not show optical objects above the plate limit within their radio contours
and are therefore considered well beyond the clusters on which they are
projected. This means that these sources are probably powerful radio
galaxies of the FR\,{\sc II} class (Fanaroff \& Riley 1974), which must await 
more sensitive optical observations to locate their optical identifications 
and measure their redshifts.

The frame for A3126\_13/16 is particularly interesting and discussed in detail
below. 

In the following we give detailed notes on those individual optical
identifications marked with an asterisk in column~10 of Table~3.
As is expanded in Section~3.5, we calculate the probability (P) of a chance 
projection based on the galaxy and star densities of Tyson \& Jarvis (1979).
P is the probability of an optical object of a given class and apparent 
blue magnitude being found at random within a circular area of
radius equal to its angular distance from the centroid of the radio
identification, as given in column~5 of Table~3. The value of P for our
chosen identification is given in parentheses after the source name
in the following notes, except for the very bright unambiguous cases such 
as A3126\_9. 

\noindent
{\bf A2746\_2a/b}.~(P=0.03)~ The radio morphology suggests a possible 
wide-angle tailed (WAT) radio galaxy with the host~G near the radio centroid. 

\noindent
{\bf A2746\_8}.~(P=0.0005)~ The St object near the radio centroid is accepted as the 
identification, but the fainter G $\sim$ 4$''$ to the east, with P=0.008, may be
an alternative identification.

\noindent
{\bf A2837\_1}.~(P=0.0002)~ Fig.~3 shows an unambiguous identification with a G near the source centroid.

\noindent
{\bf A2837\_6}.~(P=0.0008)~ The most likely identification is the G near the 
source centroid, but a fainter G (P=0.03) near the radio axis 10$''$ to the NE 
may be the host of an unresolved narrow-angle tailed (NAT) radio galaxy.

\noindent
{\bf A2837\_8}.~ Although no identification is listed, Fig.~3 shows a G 
13$''$ to the south of the radio centroid (P=0.08) that may be the host 
of an unresolved WAT. 

\noindent
{\bf A2837\_9}.~(P=0.002)~ A faint G near the radio centroid is not clear in the 
$R$-band image of Fig.~3, but is verified by the B and IR images.
The two brighter G's to the NE and NW of the
centroid could be hosts to either an unresolved NAT (P=0.002) or WAT
(P=0.002) respectively; they are probably cluster galaxies.

\noindent
{\bf A2837\_10}.~(P=0.0005)~ The faint G in Fig.~3, 1$''$ to the NNW of the 
radio centroid, is confirmed on UKST-B and ESO-R images.

\noindent
{\bf A2837\_11}.~(P=0.006)~ The bright G, 8$''$ to the ENE, could be a 
cluster member. and is accepted as the identification.
The faint object near the radio centroid appears only on the
2nd-epoch Sky Survey B-image, but not on the 1st-epoch one, and is therefore
probably a plate defect. 

\noindent
{\bf A2837\_14}.~(P=0.005)~ Figure~3 shows a  G, 4$''$ to the NE of the
radio centroid which we accept as the identification.  The object is
not bright enough for cluster membership. The faint object $\sim$15$''$
to the west has a high probability of projection.

\noindent
{\bf A3126\_2}.~ The St  7$''$ to the south of the radio centroid has P=0.02, 
but its position relative to the radio centroid and major axis does not support 
its status as an identification.

\noindent
{\bf A3126\_9}.~ The bright G is a secure identification for this extended 
source. It is a cluster member by its brightness.

\noindent
{\bf A3126\_10a/b}.~ The well-defined double source is probably a distant 
FR\,{\sc II} radio galaxy.

\noindent
{\bf A3126\_13/16}.~ Here we see two separate sources, one extended (\#13, NW) 
and the other (\#16, SE) much more compact; both are identified 
with bright elliptical galaxies, in fact the two brightest are 
spectroscopically confirmed cluster members, which are projected 
onto the cluster at angular distances of 1.0$'$ and 0.2$'$ from 
the X-ray centroid respectively. The compact SE source (\#16) coincides 
with the second-brightest cluster galaxy and is closer in projection 
to the cluster's X-ray centroid.  Given the positional uncertainty 
of $\sim1'$ of the REFLEX position for an extended source like
this (Edge 2006), A3216\_16 and its associated galaxy could well be at the
dynamical centre of the cluster. There is good evidence from the
digitized image of the 2-nd epoch UKST Red Survey for a halo surrounding
them. Another unusual feature of this pair of radio sources is the marked
difference in their radio spectra. The compact SE source has a fairly
flat spectrum ($\alpha=-0.27\pm$0.54), while the extended NW source has
a more normal spectral index of $-$0.90. The flat spectrum and compact
nature of the SE source suggest that this elliptical is dominated by an
AGN , while the NW source is possibly a partially resolved NAT whose 
projected tails extend almost to the cluster's X-ray centroid.

\noindent
{\bf A3126\_25}.~(P=0.0013)~ The comparatively bright St object, 6$''$ to 
the NNW of the radio centroid, is accepted as the identification for 
this extended source.

\noindent
{\bf A3126\_27}.~(P=0.0003)~ The faint G near the centroid of this extended 
radio source is accepted as the most likely identification. The St object 22$''$ to
the NNW of the centroid has P=0.23 and is not associated with the radio source.

\noindent
{\bf A3126\_28}.~(P=0.005)~ The comparatively bright G, 5$''$ to the SSW of 
the radio centroid, is accepted as the identification.

\noindent
{\bf A3216\_2}.~(P=0.00003)~ The bright elliptical~G near the radio centroid is accepted
as the identification, and considered a cluster member.

\noindent
{\bf A3216\_6a/b}.~(P=0.00009)~ The comparatively bright~G near the radio centroid is
accepted as the identification, and considered a cluster member. The fainter St object, 
7$''$ to the NNE, may be a projected object with P=0.023.

\noindent
{\bf A3216\_7}.~ The extended source is probably an unresolved NAT; the
host G is 16$''$ to the NE of the radio centroid along the major axis. Its
probability of projection is P=0.06 with respect to the radio centroid,
but its proximity to the major axis makes the identification more secure.

\noindent
{\bf A3216\_9}.~(P=0.00006)~ The bright G near the radio centroid is the favoured 
identification, but the other three G's within the central two contours are bright 
enough to be cluster members; they have values of P between 0.01 and 0.00007.

\noindent
{\bf A3216\_10}.~(P=0.004)~ Although the G is displaced 9$''$ from the radio centroid
along the major axis, it is accepted as the identification for this radio source. It
is bright enough for cluster membership.

\noindent
{\bf A3216\_14}.~ A bright St object (m$_R$=13.7) is 4.4$''$ from the radio centroid of
this unresolved source, but its accurate radio coordinates suggest that
the objects are not associated. There is no overlay in Fig.~3.

\noindent
{\bf A3230\_1}.~ Figure~3 shows four optical objects within the radio contours: three G's 
and one St (the fainter of the two images to the SW). All four objects
within the contours have a low probability of projection with respect to
the radio centroid (P$<$0.01) but only the bright E/S0~galaxy $\sim20''$ to 
the south east of the radio peak, and the brighter of the two fainter 
galaxies (7$''$ south of the radio peak) are bright enough to be cluster members.
The bright E/S0 SE of the radio centroid is accepted as the most likely identification,
as it is positioned along the major axis of this extended source.

\noindent
{\bf A3230\_4}.~(P=0.0002)~ The bright elliptical G near the radio centroid is accepted
as the identification for this extended source; it is possibly the host galaxy of
an unresolved WAT. Despite the clear elongation of the radio contours,
a gaussian fit was unable to yield a deconvolved size.

\noindent
{\bf A3230\_9}.~(P=0.02)~ The faint St identification for this extended source is 
displaced 6$''$ along the major axis to the NNW of the radio centroid. It is 
tabulated in the UK-R and UK-B SuperCOSMOS data.

\noindent
{\bf A3230\_11}.~(P=0.00004)~ There are two bright G's at 1$''$ and 4$''$ to the south 
of the radio centroid, but the brighter one nearer the radio centroid is accepted as 
the identification. The other galaxy is also bright enough for cluster membership. 
There is no overlay in Fig.~3.

\noindent
{\bf A3827\_2}.~ Fig.~3 shows no clear identification for this slightly extended source. 
It is possibly a distant FR\,{\sc II}.

\noindent
{\bf A3827\_3}.~(P=0.0024)~ The St object, close to the radio centroid, is accepted as
the identification. The faint G, 24$''$ along the major axis to the SSW of the radio
centroid, is probably a projected object with P=0.2; this G is too
faint to be a cluster member.

\noindent
{\bf A3827\_4}.~ The bright elliptical G near the radio centroid is clearly
the identification. The distorted nature of the outer contour to the SSE may be due
to the close negative contour. The source is clearly extended in 
PA$\sim$60$^{\circ}$, but a deconvolved size was not yielded in a gaussian fit.

\noindent
{\bf A3827\_8}.~(P=0.002)~ The bright elliptical G to the NNW of the radio centroid
of this extended source is accepted as the identification. The relatively bright St near 
the SSW outer contour may be a projected object with P=0.10.

\noindent
{\bf A3827\_11a/b}.~ The two St objects within the contours of the double
radio source are likely to be projections with P=0.21 and P=0.24
respectively. This is probably a distant FR\,{\sc II} radio galaxy.

\noindent
{\bf A3827\_14}.~ Within the radio contours are an St (24$''$@78$^{\circ}$) and a 
G (12$''$@203$^{\circ}$). Their projection probabilities are  P=0.26  and P=0.08
respectively. The G is not bright enough for cluster membership and
neither object is an acceptable identification.

\noindent
{\bf A3827\_21}.~(P=0.009)~ The faint G displaced 5$''$ along the major axis to the NNE of
the radio centroid is the accepted identification.  We discard the bright galaxy 
16$''$ to the SSE of the radio centroid (2MASX~J22015849$-$5943068) as the 
identification for its separation from the radio centroid and for its significant
displacement from the major axis of the radio source.

\noindent
{\bf A3827\_24a/b}.~ The Sp is unlikely to be a projected object whether
identified with component~b (P=0.002) or associated with the centroid
of the double (P=0.035). The identification in Table~3 is with component~b, and is
bright enough to be a cluster spiral.

\noindent
{\bf A3827\_26}.~ The faint objects near the outer contour of this extended
radio source to the east and north of the radio centroid are likely to
be projections with P=0.54 and 0.50 respectively.

\noindent
{\bf A3836\_9a/b}.~ Neither of the components of this apparent double are
resolved at 2.38~GHz and no identification  can be detected. If it is a real double,
it is probably a distant FR\,{\sc II} radio galaxy.

\noindent
{\bf A3836\_17}.~ The faint St object to the north of the radio centroid is
unlikely to be a projection (P=0.02), but its 7$''$  separation from the
centroid of the radio source and its offset from the major axis are not
consistent with an identification.

\noindent
{\bf A3836\_19a/b}.~(P=0.11)~ There is a possibility that the faint G between
the components is a projected object, but its proximity to the major
axis of the double improves the security of the identification. It is not bright
enough for a cluster elliptical.

\subsection{The physical reality of the suggested identifications} \label{reality}  

In this paper we are mainly interested in the reality of the galaxy
identifications, particularly those that have been identified as cluster
members by the process described in Section~\ref{optIDs}.

We have used the galaxy densities listed in Table~1 of Tyson \& Jarvis
(1979) to derive the probability that a galaxy will be projected
by chance onto the area surrounding a radio source in our list. Tyson
and Jarvis's galaxy densities strictly apply to the north galactic
pole. Our clusters have values of galactic latitude between $b=-37^{\circ}$
and $-51^{\circ}$, so that they can be regarded as high-latitude objects,
and we  assume that there is not a significant difference between galaxy
densities in the north and south Galactic hemispheres.

Firstly, looking at the 55 radio sources in Table~3 identified with
galaxies having listed blue magnitudes between 15.0 and 22.0, we compute that
the total area defined by the circles with radii equal to their 
listed angular offsets is 6.53$\times10^{-4}$~deg$^2$.  In this area, 
Tyson \& Jarvis give a density of 2,453 galaxies~deg$^{-2}$.  Thus we can 
expect approximately two out of the 55 coincidences by chance projection.

If we confine our attention to the 32 identified cluster galaxies in Table~3
in the complete sample, with blue magnitudes between 15.0 and 21.0, a
similar calculation yields a chance projection of 0.2 galaxies.

We therefore conclude that all the identifications in the complete sample
are highly likely to be physically associated.

\subsection{The physical reality of the double sources}  \label{physdble}  

There are seven sources in Tables~2 and 3 that are listed as double
sources. What is the probability that one of the components is a
field source that is projected on the area surrounding the other\,? The
probability of chance projection increases as the flux density decreases
and the circular area defined by the radius vector between the components
increases.

Using the  log\,N\,-\,log\,S$_{1.4}$ counts at 1.4~GHz by Wall (1994), we
computed for each double the number of sources with the flux density of
the weaker component that fall by chance in  the area of a circle with
radius equal to the angular spacing between the components. Taking into
account the range of flux densities and angular spacings in the seven
doubles, the probability of chance projection is $\le$0.002. Therefore,
we can be confident that these are highly likely to be physical double
sources rather than chance projections.

\subsection{Source identification statistics}  

Table~4 provides a summary of the types of optical object identified
with the 149 sources listed in Tables~2 and 3. These are all the sources
that can be detected at 1.38~GHz within the FWHM  circle of the ATCA
primary beams.

\setcounter{table}{3}
\begin{table}
\caption{ \noindent Identification statistics for sources in our 7 clusters
}   
\begin{tabular}{rrrrr}
\hline
Cluster &  Non-cluster & Star-like & Blank   & Total    \\
galaxies    & galaxies  &  objects  &  fields  &    \\
\hline
  N=32      & N=21     &  N=14      &  N=82     & N=149  \\
  22~\%     & 14~\%    &   9~\%  &  55~\% & 100~\%     \\
\hline
\end{tabular}
\end{table}

Only for seven of the nine galaxies with measured redshifts in Table~3 are their
redshifts commensurate with those of the clusters on which they are 
projected. The remaining 25~galaxies are allocated to the clusters on the basis 
of their absolute $R$-band magnitudes as explained in Section~\ref{optIDs}.

Very few of the 21 galaxies in the second column of Table~4 are bright
enough to be members of the cluster on which they are projected, and so
are named ``non-cluster galaxies''. There are three spirals included, two
of which possess measured redshifts, which place them in the foreground;
the third spiral in this group may be a cluster member. All 14 star-like
identifications (St) in Table~4 are fainter than $m_R$=18.1, making it difficult
to distinguish between AGN, compact ellipticals and, less likely,
real stars.

Adding the number of identifications in the first three columns of Table~4, 
we find the identification rate to be 45\%. Prandoni et al.\ (2001) searched a 
3~deg$^2$ area of their ATESP Survey for identifications with optical objects detected
in the ESO Imaging Survey (EIS) by Nonino et al.\ (1999). Confining our
attention to Prandoni et al's radio sources with 1.4-GHz flux densities
$\ge$1\,mJy, we see in their Table~2 that their identification rate for all
optical morphologies was 67\%. Our lower rate of 45\% is most likely due to
their access to a more sensitive optical survey; had a similar survey been
available to us, we would have found significantly fewer blank fields.

\section{Derived parameters} \label{derpar}    

In this section we shall be comparing the derived spectral indices, radio
powers and absolute red magnitudes of the 32 identified cluster sources in
Table~3 with a similar set of parameters published by Slee et al.\ (1998)
and derived from VLA observations in scaled C and D arrays at 1.5 and
4.9~GHz respectively. Their's was a radio-selected sample of 28 Abell
clusters, in which an earlier lower-resolution survey at 80 and 160~MHz
by Slee \& Siegman (1983) had shown a steep-spectrum radio source within a
few arcmin of the Abell cluster centre. The VLA survey of these clusters
found 59 identified cluster radio galaxies out to their limiting angular
radius of 0.5\,$R_A$ at 1.5~GHz, but fourteen of them could not be detected
at 4.9~GHz due to both their generally steep spectra and the fact that
they fell outside the FWHM primary beamwidth of the VLA dishes at 4.9~GHz.

\subsection{The radio spectra}   \label{radiospectra}   

Using the flux densities measured by us, as well as literature data
from other surveys (extracted from CATS, cats.sao.ru), we plotted radio
continuum spectra for all sources. For those with flux densities in
three or more (at most five) frequency bands, we visually classified
their shape into three classes, as listed in column~(13) of Table~2:
C$-$: convex spectrum (steepening with increasing frequency); C$+$:
concave spectrum (flattening with increasing frequency);  Cpx: complex,
including a few sources with relative minima or maxima; we do not
make any mention in this column if the spectrum is consistent with a
straight power law. The spectral indices quoted in columns~(11) and (12)
are always from straight power-law fits.  For a few sources (A2837\_10,
A3126\_11, A3827\_5, A3836\_18) our 1.38-GHz flux causes a pronounced
minimum in their spectrum. We searched for instrumental reasons or
reduction problems to explain these, but could not find any.

In order to check our flux densities for the deleterious effects
of ``CLEAN bias'', we examined some suggestions by Condon et al.\ (1998) 
and Prandoni et al.\ (2000) for keeping this bias to a minimum. It is 
clear that one needs to keep sidelobes of field sources to a minimum
and to terminate the cleaning process while its residual rms  is still
well above the the theoretical system noise. In our analysis we had
used uniform weighting to form the dirty maps and we had terminated the
de-convolution at a cut-off equal to 4 times the theoretical rms. In
order to estimate the magnitude of the bias, we also cleaned the dirty
maps down to a cut-off equal to the theoretical rms, and we compared
the peak flux densities at cut-offs of 4 and 1 times the theoretical
noise for sources with 1.38-GHz flux densities $\le$2.5\,mJy in Table~2,
resulting in mean values of (peak$_{\rm 4\,rms}$/peak$_{\rm 1\,rms}$) 
that varied between 1.06$\pm$0.04 for the three self-calibrated maps to 
1.12$\pm$0.12 for the remaining four maps, where the errors quoted 
are standard deviations.
Therefore, we can confirm that CLEAN bias is present in the lower cut-off
cleaned maps, but is probably negligible at the 4 $\times$ rms cut-offs that
were used for the flux densities in Table~2, and would be obscured by
the errors in measuring these low flux densities.

We also point out that three of the four anomalously low flux
densities were found in fields that were not subjected to
self-calibration, and the uv coverage was not unusually poor in these
seven fields south of $-51^{\circ}$ with 7--9 half-hour integrations
over the full range of hour-angles.

\begin{figure*}
\psfig{file=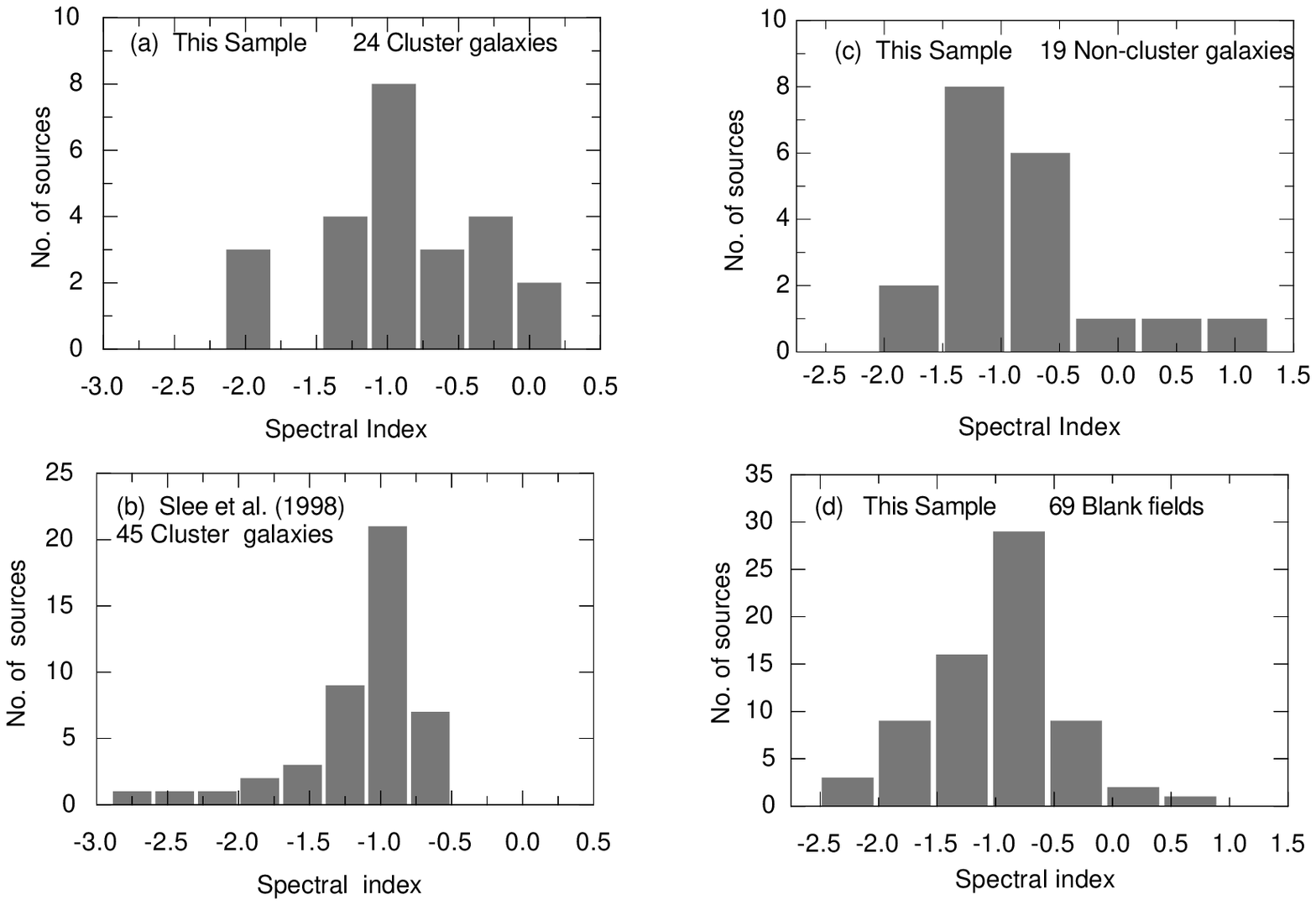,width=17.5cm,angle=0}
\caption{ \noindent Spectral index distributions for (a) 
cluster member radio galaxies in the present X-ray selected cluster sample, of 
which only 24 of 32 have spectral indices; 
(b) the radio-selected sample of Slee et al.\ (1998),
for which 14 of the 59 sources do not have spectral indices;
c) the sources in the present sample whose optical identifications were 
too faint to be members of the seven clusters in the present sample, and 
(d) for those 69 of the 82 unidentified sources in the present sample
that have spectral indices. 
Analysis of these distributions is given in Section~4.1.
}
\end{figure*}

\begin{figure*}
\psfig{file=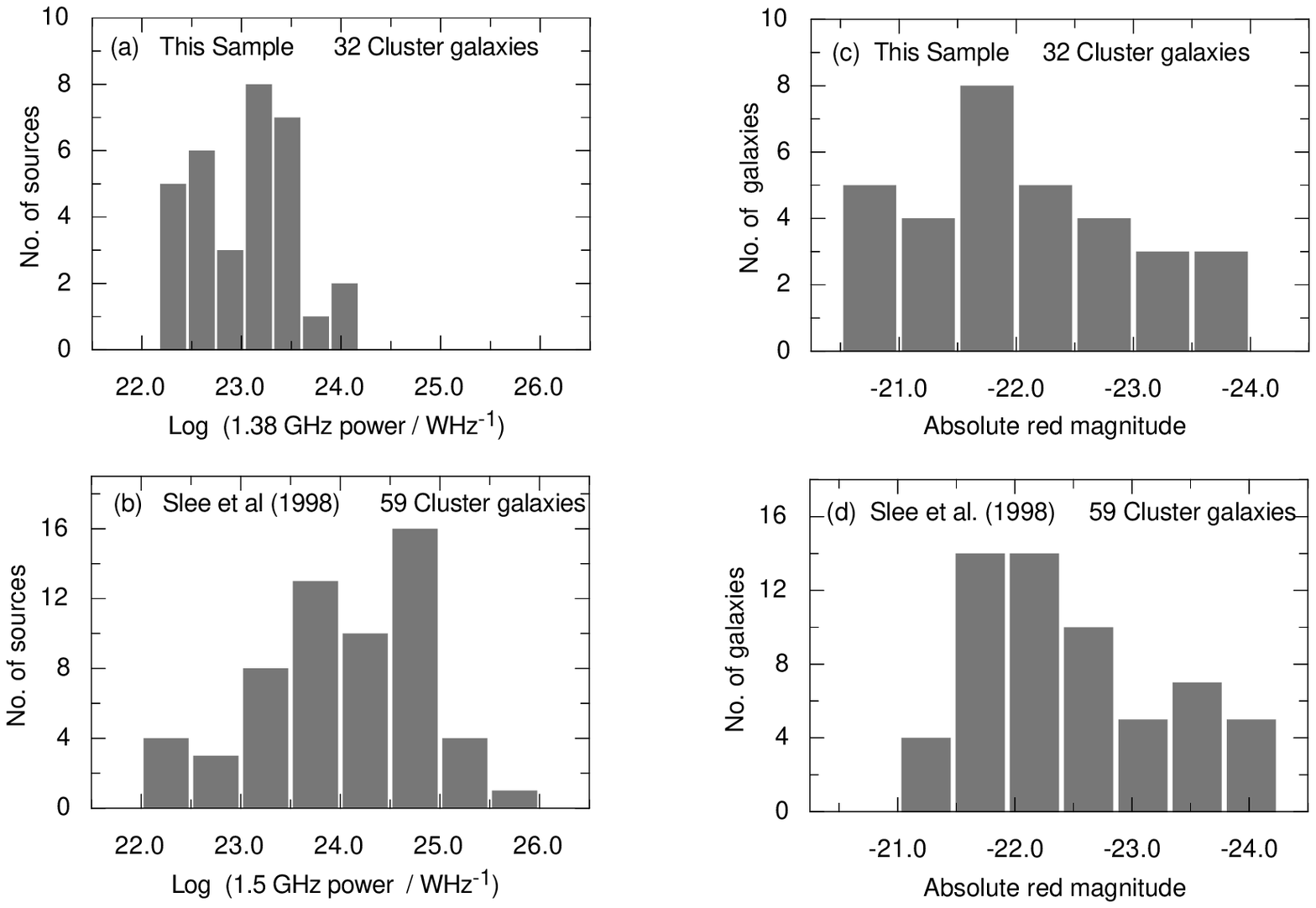,width=17.5cm,angle=0}
\caption{ \noindent 
Distributions of emitted radio power for the present X-ray
selected sample~(a) and the radio-selected sample of Slee et al.\ (1998)
(b). The corresponding distributions for absolute $R$-band magnitude are
depicted in panels (c) and (d). The significance of the differences in
these distributions is discussed in Section 4.2.
}
\end{figure*}

Figure~4 is a mosaic of four histograms of the spectral indices for a variety
of identification classes; separate plots are shown for radio sources
identified with cluster galaxies, non-cluster galaxies and optically
empty fields.  The median spectral indices are $-$0.87 (Fig.~4a), 
$-$1.02 (Fig.~4b), $-$0.96 (Fig.~4c), and $-$0.93 (Fig.~4d).
Kolmogorov-Smirnov ranking tests on the data contributing to these pairs of 
histograms indicate that (a) and (b) are significantly different ($P$=0.03); 
this difference is clear from an inspection of Figure~4, with the identified
cluster sources from the X-ray sample~(a) possessing a bias towards
flatter spectrum sources, which is reflected in their median spectral
indices. The only other marginally significant difference ($P$=0.08) is between
the pair of histograms~(b) and (d).

\subsection{Radio power and absolute $R$-band magnitude}  \label{radpowMR}  

Figure~5 is a montage of four histograms depicting the radio powers and
absolute $R$-band magnitudes of two samples of identified cluster galaxies. The
first sample comes from the present seven X-ray-selected clusters, the
second sample is from the radio-selected clusters from Slee et al.\ (1998). 

The X-ray sample~(a) has a median 1.38-GHz power of log(P/W\,Hz$^{-1}$)=23.08, 
which is an order of magnitude lower than that of the radio sample~(b) with a 
median power of log(P/W\,Hz$^{-1}$)=24.07. This significant
difference is amply confirmed by a Kolmogorov-Smirnov test on their
data-distributions, with the probability that they originate from the same
parent populations of $P\le$0.005. This result is not unexpected, as
the clusters that contribute to sample~(b) were selected because their
optical centres were close to strong radio sources.

The corresponding distributions of absolute $R$-band magnitude for the galaxies
identified with radio sources are shown in Figure~5, panels (c) and (d), and are
not significantly different, with the probability that they originate
in the same parent population of $P$=0.26; the median values of $M_R$ for
(c) and (d) are $-$21.93 and $-$22.34 respectively. This relatively small
difference  suggests that the radio and optical power outputs of cluster
galaxies are not closely related. This is already obvious from the fact
that most of the relatively bright galaxies in both samples of clusters
are not identified with radio sources at the 1~mJy  level  at 1.38 and
2.38~GHz, a conclusion that will be made clearer when the fractional
radio luminosity function is discussed in Section~\ref{RLF}.

\subsection{Radial variation in source density}  \label{radvar}  

The degree of concentration of cluster radio galaxies towards the cluster
centre is an important parameter that is probably influenced by the
evolution of their host galaxies and the radial variation in intracluster
gas density. Slee et al.\ (1998) showed the radial variation in projected
source density as function of the logarithm of the projected cluster-centric
distance, in units of an Abell radius (r/$R_A$), in their Figure~4. Here, 
we compare that result with a similar analysis of the present X-ray sample.

We counted cluster sources in equal increments of angular radius
centred on the seven X-ray cluster centroids out to the Abell radius of
2~Mpc. Adding the numbers of sources in each angular increment produces the
histogram in the top left panel of Figure~6. Adjusting the numbers for the
projected areas of the annuli (in units of Mpc$^2$), we show the projected
source density as a function of log\,r/$R_A$ in panel~(b) of Figure~6.

Panels~(c) and (d) of Figure~6 show the corresponding results from
the radio sample of Slee et al.\ (1998). Due to the larger number of cluster
sources available (59 versus 32 for the X-ray selected sample), 
we were able to use a smaller increment in r/$R_A$. The slope of
the fitted line in this sample is $-1.55\pm$0.06.

Other determinations of the concentration of sources towards the cluster
centres have been made by Mills \& Hoskins (1977), Slee et al.\ (1983),
Robertson \& Roach (1990), \cite{andand90}, Unewisse (1993), and 
Ledlow \& Owen (1995a).  These studies utilized surveys with widely
varying sensitivities and angular resolutions, but all of them (except
the one in Fig.~1d of Andernach \& Andreazza 1990) derive radial plots
with slopes similar to or steeper than those in panels (b) and (d)
of our Figure~6.

The 1.4-GHz survey of Ledlow \& Owen (1995a) is closest in frequency,
angular resolution and sensitivity to the ATCA and VLA data plotted in panels
(b) and (d) of Figure~6. We have replotted the FR\,{\sc I} data shown
in their Figure~2A with logarithmic axes (the numerical value of their 
last bin was difficult to read with enough accuracy and so was omitted);
after fitting a straight line to their points we derive a slope of 
$-1.83\pm0.19$, which is significantly steeper than those of our Figures~6b
and 6d. Evaluating the regression equation yields a projected source
density at r/$R_A$=0.03 of 10~Mpc$^{-2}$, which is similar to our values
from panels (b) and (d), but because of the steeper slope the projected
source densities at r/$R_A$=1 are four to eight times lower, respectively.
The difference between our radial dependencies in Figure~6 and those
of Ledlow \& Owen are probably due to our samples having flux-density
cut-off levels of 1.0 and 2.0~mJy, respectively, compared with 10~mJy
for the VLA sample of Ledlow \& Owen. Our higher sensitivity results
in twenty-six of our 32 identified cluster radio sources having flux
densities below 10~mJy, and twenty of those fall outside the 0.3~$R_A$
limit of Ledlow \& Owen's complete sample. We contend that Ledlow \&
Owen's projected cluster source densities are not taking into account
a significant population of less powerful cluster radio sources.

An important task is to establish the average projected density of
background sources in order to derive the excess of cluster radio sources
with respect to the background. Our initial approach to this problem
was to plot the log of the average projected source density of the 117
``non-cluster'' sources (see Table~4) against log~r/$R_A$.  Fig.~6e
shows their distribution in r/$R_A$, while the dashed line in Fig.~6f
outlines their projected source density as a function of log(r$/R_A$);
this line appears to show the presence of another unexpected population
of cluster sources.  A second approach was to use the source statistics
of Prandoni et al.\ (2001) from their ATESP survey of 26~deg$^2$ of
the southern sky; these authors used the AT Compact Array in a similar
configuration to that of our seven-cluster survey. We counted their
radio sources with 1.4~GHz flux density $\geq$0.99 mJy (the lower limit
of our survey) to derive an average source density of 72~deg$^{-2}$; this
figure results in a projected source density of 1.59 sources Mpc$^{-2}$,
assuming H$_0$=75\,km\,s$^{-1}$\,Mpc$^{-1}$.  For comparison, our 149
sources in seven clusters result in an average projected source density
of 77~deg$^{-2}$, indicating a surprising level of agreement between two
independent surveys.  A dotted horizontal line corresponding to this
source density is shown in Fig.~6f.

We see that the dotted line passes close to most of the points in Fig.~6f, 
suggesting that our ``non-cluster'' source density is close to the background
density.  The slope of $-0.5\pm$0.12 is not convincingly different enough
from the zero slope of the ATESP background to support the existence
of another population of cluster radio sources, although there is
almost certainly a fair proportion of spiral galaxies included in our
seven clusters.  If we consider only the eighteen G and E ``non-cluster''
identifications in Table~4, we find that, if we locate them within their
relevant clusters, they would have a median $M_R=-$19.48 and a median
log\,P$_{1.38}$/W\,Hz$^{-1}$=22.91. This value of $M_R$ is about 1.5 magnitudes
lower than that found by Hummel (1981) for the median absolute red
magnitude of 280 local spirals of all classifications, and the value of
log\,P$_{1.38}$ exceeds the high-power end of his fractional radio luminosity
function. Extending this type of analysis to the 82 empty fields in
Table~4, only accentuates the difference between Hummel's spirals and
our non-cluster sample of radio galaxies. We therefore conclude that
very few of the sources contributing to the dashed line in Fig.~6f
can be spirals of the type seen in the local volume.

\begin{figure*}
\psfig{file=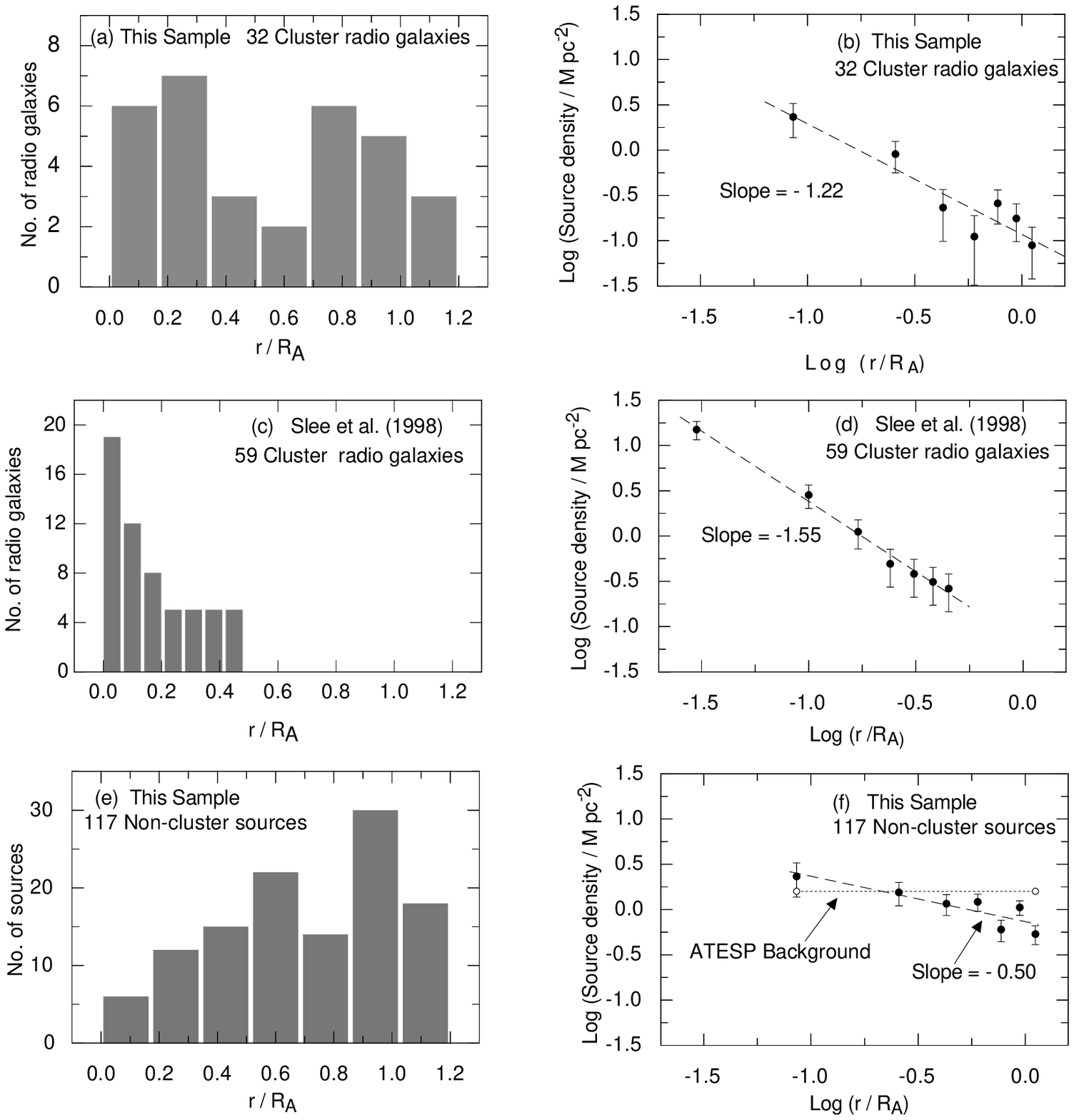,width=17.5cm,angle=0}
\caption{ \noindent 
Histograms of radio source distributions in r/$R_A$ are shown in panels~(a), 
(c) and (e). The corresponding projected source densities as a function of 
log(r/$R_A$) appear in panels~(b), (d) and (f). The dotted horizontal line 
depicts the average projected radio source density derived from the 
ATESP survey.}
\end{figure*}

\begin{figure*}
\psfig{file=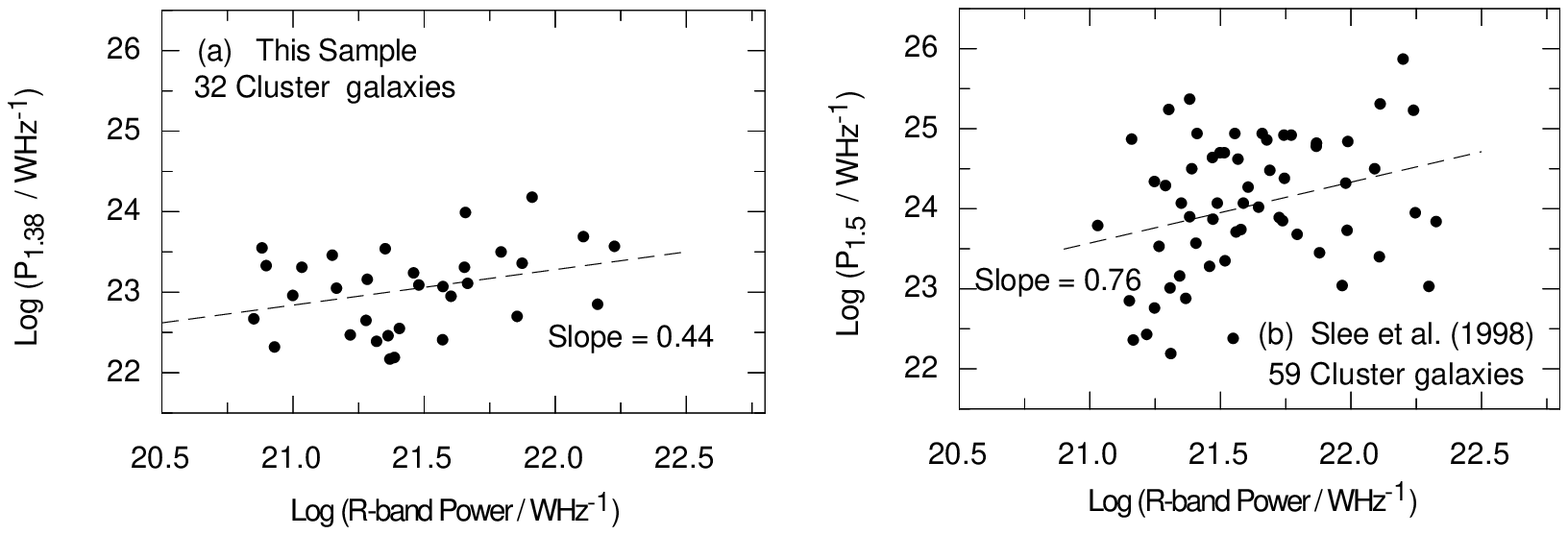,width=17.5cm,angle=0}
\caption{ \noindent 
Radio power against $R$-band power for the X-ray selected and radio samples is
plotted in panels (a) and (c) respectively. The corresponding plots of powers, 
summed over all sources of each cluster, are shown in panels (b) with 7 clusters
and (d) with 28 clusters.  Linear fits are made to the log--log data points.
See the text of Section~4.4 for the slopes of the dashed regression
lines and their significance.
}
\end{figure*}

We accept that the background source density at 1.38\,GHz is 1.59~Mpc$^{-2}$
as defined by the horizontal dotted line in Fig.~6f. Thus referring to
Fig.~6b, we find that the average projected source density near the 
cluster centres (at $r/R_A=0.03$) for ellipticals in our X-ray sample is 
5.0~times the projected source density of background radio sources. 

We can conclude this section by stating that the higher intracluster
gas density of our X-ray selected sample does not appear to significantly
affect either the central excess density of radio sources above the 
background or its variation with projected radial distance from the cluster centre.

\subsection{Correlation between radio and optical power}  \label{radoptpower}  

We have already shown in Section~\ref{radpowMR} and the histograms in Figure~5
that while the distributions in the radio powers of the X-ray selected
and radio-selected samples are very different, their distributions in
absolute $R$-band magnitude cannot be readily distinguished. This fact alone
implies that there should not be a strong correlation between radio and
optical power in either sample. In order to keep the same units on the
axes of the correlation plots we have transformed $R$-band luminosity
to $R$-band power. Here, we examine the detailed correlation between radio
and optical powers in the two samples.

Figure 7 shows the correlation between radio power and $R$-band power for
the present X-ray sample in panel~(a) and for the radio-selected sample
of Slee et al.\ (1998) in panel~(b); the former and latter samples show
reasonably well-defined correlations with significance levels of 9\%
and 3\% respectively.  The slopes of the regression lines in (a) and (b)
are 0.44$\pm$0.24 and 0.76$\pm$0.33 with a weighted mean of 0.55. Due to
the overlap in the error bars in the slopes of the samples, it is not
possible to show from the correlations in Figure~7 whether the slopes
of their regression lines differ significantly.

The observed radio-optical power correlation could be intuitively
expected if the more luminous optical galaxies inject more relativistic
plasma into the radio lobes via the jets. However, it would be na\"{i}ve
to propose that this is the only influence on the emitted radio power.
Indeed, the correlation coefficient of $r$=0.29 in the data of Figure~7b 
implies that only 8\% of the variance can be attributed to this
correlation. Additional evidence for these correlations for FR\,I radio
galaxies is provided in the plots of log\,P$_{1.4}/M_{24.5}$ in Figure~3 of 
Ledlow \& Owen (1996), although these authors do not attempt to fit
their scatter diagram with formal regression lines. As is the case for our 
data, these authors show that there is no strong relationship between radio
power and $M_R$, which they attribute to the fact that one sees radio 
sources in all stages of their evolution.

\subsection{Correlations between various radio parameters and r/$R_A$}  

\subsubsection{Correlations between radio and optical parameters and Abell radius} 

There is little firm evidence that radio spectral index, radio power
and red optical power are highly dependent on angular distance from the
cluster centre (r/$R_A$). This is true of both the present X-ray selected
sample and the radio-selected sample of Slee et al.\ (1998). We tabulate
the significance levels of the various correlations for our X-ray selected
sample below:

\begin{tabbing}
~~~1.~~~~ \= Spectral index~\, vs.~~r/$R_A$  \= ~~~~26\%  \\
~~~2. \> ~~~~ Log\,P$_{1.38}$~~~~ vs.~~r/$R_A$  \> ~~~~86\% \\
~~~3. \> ~~~~~~ Log\,P$_R$~~~~~~ vs.~~r/$R_A$  \> ~~~~10\%
\end{tabbing}

Very similar levels of significance were obtained for the corresponding
correlations in the radio selected sample of Slee et al.\ (1998). The
only correlation that even approaches a significant level of acceptance
in both these samples is that between $R$-band power and r/$R_A$, with the
higher values of $R$-band power favouring the cluster centres. The lack of
a significant correlation between spectral index and r/$R_A$ is rather
surprising in view of the fact that lower angular resolution studies
by Baldwin \& Scott (1973), McHardy (1979) and Reuter \& Andernach (1992)
have shown that radio spectra tend to be steeper closer to the cluster centre.

\subsubsection{Correlations of radio and $R$-band power with X-ray luminosity}  

There are only seven clusters in the X-ray sample and 28 clusters in the 
radio sample of Slee et al.\ 1998, and, furthermore, only sixteen of the 
latter possess published values of X-ray luminosity. The only possible method 
of examining the relationships of radio and $R$-band power with X-ray luminosity
is to sum the radio and $R$-band powers from 
the identified cluster sources and plot these sums against the the X-ray 
luminosities of the corresponding clusters. The X-ray luminosities were obtained 
from the published  ROSAT (RASS) surveys by Ebeling et al.\ (1996, 1998, 2000), 
B\"ohringer et al.\ (2000, 2004), De\,Grandi et al.\ (1999), Cruddace et al.\ (2002), 
Ledlow et al.\ (2003) and  David, Forman \& Jones (1999). The published
X-ray luminosities were corrected to H$_0$ =75\,km\,s$^{-1}$\,Mpc$^{-1}$, 
and duplicated values for the same cluster were averaged.

We found no significant correlation between the summed radio and $R$-band
powers and the corresponding X-ray luminosities for the present
X-ray selected sample, with levels of chance probability of 82\% and 60\%
respectively. The larger radio-selected sample of 16 clusters from Slee
et al.\ (1998) yielded a chance probability of 65\% for the correlation
of summed radio power, and the more significant value of 6\% for the
correlation of $R$-band power; the linear regression of summed $R$-band power 
on X-ray luminosity has a slope of 0.49.
These would be  important relationships to better establish, but it is
clear that this will need much larger samples of clusters with measured
X-ray luminosities.

\subsection{Radio luminosity functions}   \label{RLF} 

Our seven X-ray clusters provide only 32 optically identified cluster
members with which to construct a fractional radio luminosity function
(FRLF). The FRLF is intended to show what fraction of E/S0 galaxies in
the seven clusters are detected radio emitters above the flux density limit
of the survey. It is a differential luminosity function in that the
fraction of radio emitters is computed separately for each of a number
of bins of increasing radio power.

First, it is necessary to establish the total number of E/S0 galaxies that
are likely to be cluster members. Up to the present, no attempt has been
made to morphologically classify the galaxies that are present within
the Abell radius of each of our clusters in a manner similar to that of
Dressler (1980), who did so for 55 rich clusters. Digital processing
of sky survey plates can provide a reliable set of blue, red and infra-red
magnitudes and also classify the objects as galaxies or star-like. We
have used arguably the best of these digital catalogues, SuperCOSMOS,
to select E/S0 cluster galaxies for the purpose of normalizing our
bin-counts of optically identified cluster radio sources. The process
consists of the following steps:

(1) Select from Table~3 the E, cD, D and S0 galaxies within one
Abell radius of the clusters' X-ray centres, including the designated
radio/optical identifications; this involves our computing the 
absolute $R$-band magnitudes, $M_R$, of the relevant objects on the assumption 
that galaxies  within the clusters will have $M_R\le-$20.6
(as explained in Section~\ref{optIDs}). A total of 42~galaxies was selected.

(2) Find the mean value of colour index $B-R$=1.66$\pm$0.39 (standard
deviation) for these 42 galaxies using $B_J$ and $R$ from columns (3)
and (4) of Table~3.

(3) For each cluster, search the Abell area about the cluster centre in
SuperCOSMOS, selecting objects with mean class~=~1 (galaxies), $m_R\le$
magnitude for which $M_R\le-$20.6 at the redshift of the cluster, and 
$B-R\ge$1.27; the last constraint comes from (2) above.

The application of the above constraints resulted in our selecting between
101 and 263 galaxies for each of the seven clusters with a total of 1158~galaxies. We
think the great majority of them will be E/S0 galaxies because of two
reasons: spirals usually possess $M_R>-$20.5 and their $B-R$ values are
considerably less than those of ellipticals at the same redshift.

\begin{figure*}
\psfig{file=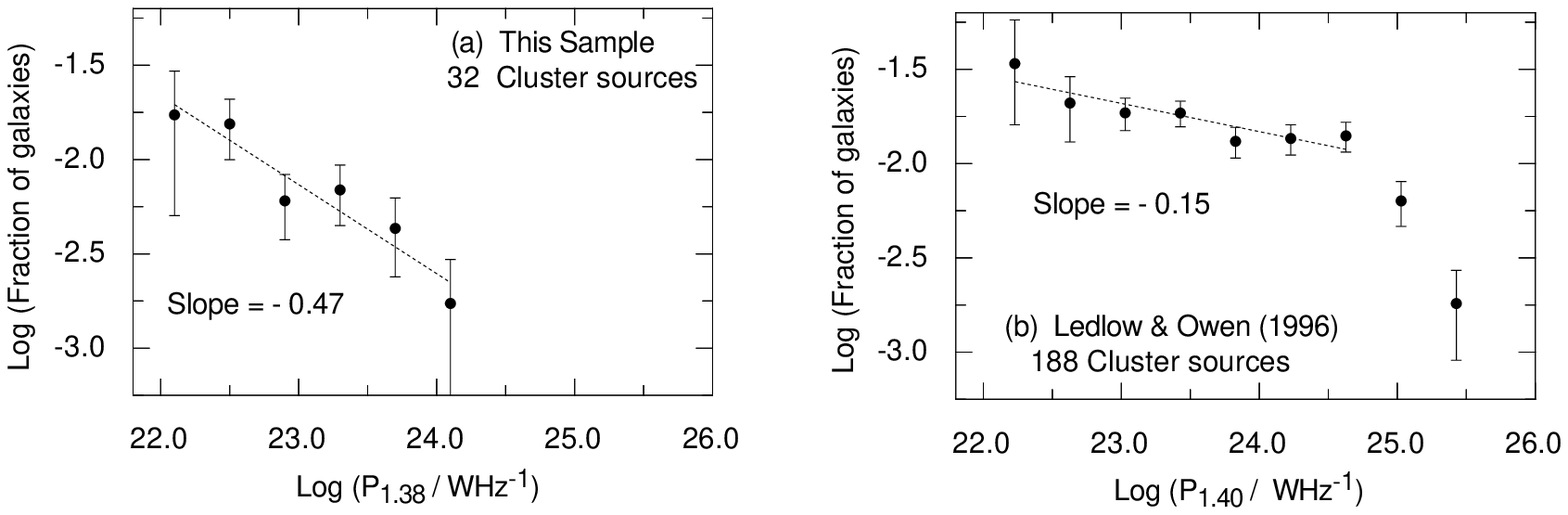,width=17.5cm,angle=0}
\caption{ \noindent 
Fractional radio luminosity functions for sources in the present X-ray
sample and in the radio-selected sample of Ledlow \& Owen (1996) are
shown in panels (a) and (b) respectively. The linear regression lines are
fitted to those parts of the data to which they seem appropriate. Error
bars determined by Poisson statistics are shown. The slopes of these
regressions and their implications are discussed in Section~\ref{RLF}.
}
\end{figure*}

The radio powers were allocated to bins, each of width 0.40 in log\,P$_{1.38}$
and the totals summed over the six bins. In this process, not all clusters
could contribute to each bin; only those clusters with redshifts low
enough to ensure that the radio identification would contribute a 1.38-GHz 
flux density greater than our sensitivity limit of 1.0~mJy could
contribute to a particular bin. Thus the first two bins with lowest
radio power could utilize only one and four clusters respectively,
while the remaining four bins utilized the radio power values from all
seven clusters. The fractions relevant to each bin were then computed by
dividing the summed numbers of radio sources in each bin by the number
of E/S0 galaxies that contribute to each bin.

The resulting fractional luminosity function for the present X-ray sample
of clusters can be seen in Figure~8a; sampling errors are proportional
to (N)$^{-0.5}$, where N is the number of radio galaxies in each bin. 
A similar FRLF was shown by Slee et al.\ (1998) in their Figure~17, but for 
comparison we prefer to plot in Figure~8b the RLF from Ledlow \& Owen (1996), 
who utilized their data from 137 radio-selected Abell clusters.
The major difference in the derivation of these plots is that the data
for Fig.~8a come from projected angular distances in the clusters out to
r/$R_A$=1.0, while Fig.~8b includes data out to only 0.30~r/$R_A$, and
clusters with redshifts $\le$0.09. The linear regressions in Figs.~8a and
8b are fitted to sections of the luminosity functions to which they
seem appropriate.

The obvious differences between Figs.~8a and 8b are in their ranges of 
radio power and in the slopes of the lines fitted to the points. 
The radio-selected sample shown in Fig.~8b contains radio galaxies that
are up to $\sim$20 times more powerful than those in the X-ray selected
clusters in Fig.~8a; this is rather surprising when one considers that
the explored cluster volume in the X-ray sample of Fig.~8a is 1.9 times
that of the radio-selected sample in Fig.~8b. One could argue that this
discrepancy is due to the much larger number of clusters in Owen \& Ledlow's
sample, and to the fact that they are exploring cluster volumes
nearer the cluster centres, but we have shown in Section~4.5.1 that
the radio power of identified cluster sources does not significantly
increase towards lower values of r/$R_A$. Ledlow \& Owen (1996) claim
that the break in their FRLF near log\,(P$_{1.4}$/W\,Hz$^{-1}$)=24.6
denotes the onset of the FR\,{\sc II} phenomenon.  It is clear that
our X-ray sample contains only the weaker FR\,{\sc I} radio galaxies,
but, like the radio-selected sample of Ledlow \& Owen in panel~(b),
the radio-selected sample of Slee et al.\ (1998) also contains three
FR\,{\sc II} radio galaxies.

The next feature of interest concerns the slopes of the regression lines
in Figures~8a and 8b. Those slopes are $-0.47\pm$0.08 and $-0.15\pm$0.03
for the X-ray and  radio samples respectively. The two plots begin with
similar values (0.017 and 0.034 respectively) of fractional luminosity 
near 10$^{22}$\,W\,Hz$^{-1}$,
but the X-ray sample~(a) falls rapidly to 0.0017 in the highest radio
power bin, while the radio sample of Ledlow \& Owen falls to only 0.014
over the same power range. The alternative radio sample of Slee et
al.\ (1998) in their Figure~17 has a similar very small slope of $-$0.06
over the FR\,{\sc I} range and then rapidly steepens to a slope of $\sim-$0.6 over 
the FR\,{\sc II} range. One might question whether the respective
luminosity functions are influenced significantly by the different ranges
of angular distance used in their compilation. It seems unlikely that
this factor has a great influence; Figure~6 in Section~4.3 shows that
the radial variation in projected source density for the X-ray sample~(Fig.~6b), 
based on sources to r/$R_A$=1, is nominally less steep than that
of the radio sample (Fig.~6d), drawn from sources out to r/$R_A$=0.50,
although the difference in their slopes is not very significant.

We conclude that this X-ray sample has a fractional luminosity
function for radio galaxies that falls much more rapidly than that of
radio-selected samples; in other words, a much smaller fraction of the
E/S0 galaxies in the X-ray sample gives rise to high-power FR\,{\sc I} radio
galaxies, and none of the cluster galaxies is host to an FR\,{\sc II} radio
source. The implications of this result will form part of the discussion.

\section{Discussion}     \label{discussion}   

Our results do not encompass a sufficiently large range of redshift,
nor are there a sufficient number of clusters, to search for evolution in
strong X-ray emitting clusters, but we can draw a few conclusions about
the effects of the cluster environment on the evolution of galaxies and
their associated radio sources. 

First, 25\% of the identified cluster radio sources possess spectral
indices more positive than $\alpha=-$0.50, which are distributed over
the full range of angular distances out to the Abell radius. Such flat
or even rising spectra are also associated with the active nuclei of
isolated E/S0 galaxies (cf.\ Slee et al.\ 1994); therefore it seems
likely that a significant proportion of radio sources in strong X-ray
clusters may be undergoing an active phase, perhaps because their black
holes are fuelled by the copious intracluster gas, although there is no
clear tendency for them to favour the inner regions of these clusters
where the gas density is highest.

Both Slee et al.\ (1998) and Ledlow \& Owen (1996) found a positive
relationship between radio and optical powers, and the effect is present,
but with lower significance, in our small X-ray sample. The radio and
optical powers are not strongly correlated for individual galaxies;
cuts through Figures~7a and 7b at constant $R$-band power shows a
large variation of at least an order of magnitude in radio power. The
interpretation of these data is very model-dependent. One can reasonably
assume that the ages of FR\,I radio sources are at least an order of
magnitude lower than that of their host galaxies, but the effect of the
host galaxy's gas content, and of the intracluster gas density, on the
evolution of an FR\,{\sc I}'s jets and lobes are largely unknown. If one
also assumes that the optical luminosity of the host galaxy is fairly
constant over the lifetime of the radio source, and that its radio power
evolves with time as the source expands into its local environment, then
the density of points along a line of constant optical power represents
the amount of time radio sources spend in that state. Unfortunately, we
do not have enough data from this small X-ray-selected sample of clusters
to construct a bivariate luminosity function, which would show the rate
at which radio power evolves at various values of optical luminosity.

The fraction of E/S0 galaxies in the optical and radio samples of Ledlow
\& Owen (1996) and Slee et al.\ (1998), respectively, that host FR\,{\sc
I} radio galaxies at a particular epoch is generally 0.013--0.025.
Ledlow \& Owen (1996) show that this fraction seems to be independent of
the environment, i.e., whether the E/S0 galaxies are in clusters or in
the general field. In addition, the integrated bivariate radio luminosity
functions of Ledlow \& Owen (1996) and Slee et al.\ (1998) indicate that
the detectable fraction increases with the optical luminosity of their
host galaxies over the complete range of FR\,{\sc I} power. If we add
to these facts that Ledlow \& Owen (1996) find from surface photometry
that the hosts of cluster FR\,I galaxies are optically indistinguishable
from radio-quiet cluster galaxies down to their sensitivity limit at 1.40
GHz of 10~mJy, then clearly the optical power output is a sufficient and
necessary condition for the production of an FR\,I radio galaxy. It then
follows that fainter  E/S0 galaxies with $M_R>-$20.5 (the limit of the
present samples) will be detectable with similar frequency at 1.38-GHz
radio powers below 10$^{22}$\,W\,Hz$^{-1}$. None of these parameters is
inconsistent with the suggestion that all E/S0 galaxies, whether in or
out of clusters, will host an FR\,{\sc I} radio galaxy at least once during
their lifetimes.

A notable difference between our X-ray sample and the radio/optical
samples is our failure to detect a single FR\,{\sc II} cluster radio
galaxy.  FR\,{\sc II} sources are found with equal frequency both in clusters
and in the general field (Ledlow \& Owen, 1996) and are characterized
by two parameters: (1)~log\,(P$_{1.4}$/W\,Hz$^{-1})\ge$24.0, 
(2)~edge-brightened lobes, caused by the interaction of the jets with the
surrounding gas. The FR\,{\sc II} phenomenon is not confined to the brightest
galaxies, but is found also in ellipticals down to a luminosity of
$M_R=-$21.0, which encompasses practically all the identified cluster
galaxies in our X-ray sample and the radio samples of Slee et al.\
(1998) and Ledlow \& Owen (1996). If FR\,{\sc II} sources had been contained
within our X-ray selected sample, we had the necessary angular resolution,
especially at 2.38~GHz, to resolve their outer hot spots from the general
lobe structure to confirm the evidence provided by the steepening of
the RLF in the samples of Ledlow \& Owen (1996) and Slee et al.\ (1998).
Ledlow \& Owen (1996) listed 18 FR\,{\sc II} radio galaxies in their complete
sample, which surveyed a projected sky area of 155\,Mpc$^2$. Slee et al.\
(1998), in their sample out to r/R$_A$=0.5, surveyed a projected area
of 71\,Mpc$^2$. Therefore, if the radio sources in these two samples had
been drawn from the same populations, we would have expected to detect
(71/155)$\times$18=8.2 FR\,{\sc II}'s and we detected nine FR\,{\sc II}'s. 
The present X-ray sample surveyed a projected area of 88\,Mpc$^2$, so we 
should have expected to detect (88/155)$\times$18=10.2 FR\,{\sc II}'s, 
but we found none.

It seems probable that the higher luminosity X-ray clusters are deficient
in FR\,{\sc II} radio galaxies, although no apparent reason is evident,
as there is more than sufficient intracluster gas with which the jets
can interact to cause edge-brightening. Perhaps the higher gas density
inhibits the formation of long jets. If this is a valid phenomenon among
clusters, then the clusters containing FR\,{\sc II} sources should possess
significantly lower values of $L_X$ than those containing only FR\,{\sc
I}s. We have searched for this effect in the radio samples of Slee et al.\
(1998). If we consider the median values of $L_X$ in their radio sample
out to 0.50~r/$R_A$, for the seven clusters containing FR\,{\sc II}
sources and the 21~clusters containing only FR\,Is, we find (including
upper limits of 0.2$\times10^{37}$\,W for the 12~clusters that were
well surveyed with ROSAT in the 1RXS catalogue, but not detected)
that the medians for the FR\,{\sc II} and FR\,{\sc I} clusters are
$<$0.2 and 0.81$\times10^{37}$\,W respectively. This result suggests
that low values of $L_X$ are associated with FR\,{\sc II} sources and
intermediate values of $L_X$ with FR\,{\sc I}s. This conclusion is
reinforced when we compare the above results with the median value
of $L_X=2.31\times10^{37}$\,W in the present X-ray selected sample,
in which no FR\,{\sc II}s are found. There is, however, an exception
to this trend in Slee et al's radio sample, in which an FR\,{\sc II}
is present in one cluster with $L_X=6.48\times10^{37}$\,W.  Clearly,
larger samples of X-ray selected clusters and their associated sources
need to be examined before a firm decision can be made.

The lack of FR\,{\sc II} radio galaxies in our present sample is
consistent with the scenario advanced by Hardcastle et al.\ (2007) for
the powering of jets in low and high-powered radio galaxies. The FR\,{\sc
I}'s are seen as low-excitation objects, which lack the narrow-line
optical emission expected from conventional AGN. In a diagram of K-band
vs.\ 151-MHz luminosity (Fig.~1 of Hardcastle et al.), these galaxies
congregate in the area occupied by low to medium values of K-band and
radio luminosity. The FR\,{\sc II}'s, on the other hand, show the more
typical parameters of AGN with high-excitation, narrow-band emission and
occupation of the medium-to-high area of the K-band -- radio luminosity
diagram.

Hardcastle et al.\ (2007) present a convincing argument that the
jets of FR\,{\sc I}'s are powered by accretion on to a black hole of hot
ionized gas from its surrounding medium. These so-called hot-mode
systems require a plentiful supply of hot gas and a massive central
black hole. By contrast, the jets of high-excitation galaxies (FR\,{\sc II}'s)
are maintained directly by cold-mode accretion, and neither
need a rich, hot-gas environment nor to be at the bottom of a deep
potential well, allowing galaxy mergers (ellipticals with spirals)
to take place.

It therefore seems likely that galaxy clusters with strong X-ray emission,
implying the presence of a dense, hot ionized gas, will possess the
required features to fuel the jets of FR\,{\sc I}'s. If this model is true,
then there should be an X-ray luminosity below which the jets of FR\,{\sc I}'s
are not so readily fueled. This means that the radio-selected samples
of galaxy clusters will contain a mixture of FR\,{\sc I}'s and FR\,{\sc II}'s, 
as is reported in the present paper.

In a survey based on ROSAT-PSPC data of a large optically selected
sample of clusters with ACO richness class R$\ge$2, David et al.\ (1999)
showed that the dominant correlation was between $L_X$ and ACO galaxy
count (i.e., richness); for a given optical richness there is also a
correlation between $L_X$ and Bautz-Morgan type, with BM\,I clusters
being the most luminous. No obvious relationships of this kind are seen
in our present sample of clusters, which encompasses Bautz-Morgan types
from I to III, ACO richness classes from 0 to 2, and with its X-ray
luminosities restricted to a small range of 2.65:1. It can be said,
however, that the median X-ray luminosity of 2.32$\times10^{37}$\,W
is comparatively high for a sample that has such diverse values of
Bautz-Morgan type and optical richness.

\section*{Acknowledgements}

The Compact Array is part of the Australia Telescope, which is funded
by the Commonwealth of Australia for operation as a National Facility
managed by CSIRO. 
The MOST is operated by the School of Physics with the support of the
Australian Research Council and the University of Sydney. 
We acknowledge our extensive use of the SuperCOSMOS database, and we
also used the CATS database maintained at the Special Astrophysical
Observatory, Russia.  
HA acknowledges financial support from CONACyT grant 40094-F, as well as
help with preparing radio spectra from A.~Maldonado and S.~Jeyakumar.
We are indebted to B.~Komberg for valuable discussion and A.\,C.~Edge for
useful advice. We thank the anonymous referee for a careful reading
and various suggestions that helped improve the text.



\begin{thebibliography}{}

\bibitem[Abell et al.\ (1989)]{aco89} Abell, G. O., Corwin~Jr., H. G. \& Olowin, R. P. 1989, ApJS, 70,1
\bibitem[Andernach \& Andreazza (1990)]{andand90} Andernach, H. \& Andreazza, C. M. 1990, Rev.\ Mex.\ Astron.\ Astrof., 21, 136
\bibitem{}Baldwin, J. E., \& Scott, P. F. 1973, MNRAS, 165, 259
\bibitem{}Bock, D. C.-J., Large, M. I., \& Sadler, E. M. 1999, AJ, 117, 1578
\bibitem{}B\"ohringer, H., Voges, W., Huchra, J.~P., et al.\ 2000, 129, 435
\bibitem{}B\"ohringer, H., Schuecker, P., Guzzo, L., et al.,
    2004, A\&A, 425, 367
\bibitem{} Clarke, J.~N., Little, A.~G., \& Mills, B.~Y. 1976, AuJPA, 40, 1
\bibitem{} Colless, M., \& Hewett, P., 1987, MNRAS, 224, 453
\bibitem{} Condon, J. J., Cotton, W. D., Greisen, E.~W., et al.\ 1998, AJ, 115, 1693
\bibitem{} Cram, L., \& Ye, T. 1995, AuJPh, 48, 113
\bibitem{} Cruddace, R., Voges, W., {B\"ohringer}, H., et al.\ 2002, ApJS, 140, 239
\bibitem{} David, L.P., Forman, W., \& Jones, C. 1999, ApJ, 519, 533
\bibitem{} De\,Grandi, S., B\"ohringer, H., Guzzo, L., et al.
     1999, ApJ, 514, 148
\bibitem{} De\,Grijp, M. H. K., Keel, W.~C., Miley, G.~K., Goudfrooij, P. \& Lub, J. 1992, A\&AS, 96, 389
\bibitem{} Dressler, A. 1980, ApJS, 42, 565
\bibitem{} Ebeling, H., Voges, W., B\"ohringer, H., et al.\ 1996, MNRAS, 281, 799
\bibitem{} Ebeling, H., Edge, A.~C., B\"ohringer, H., et al.\ 1998, MNRAS, 301, 881
\bibitem{} Ebeling, H., Edge, A. C.; Allen, S. W., et al.\ 2000, MNRAS, 318, 333
\bibitem{} Edge, A. C. 2006, priv.\ comm.
\bibitem{} Fanaroff, B. L., \& Riley, J. M. 1974, MNRAS 167, P31
\bibitem{} Frei, Z., \& Gunn, J. E. 1994, AJ, 108, 147
\bibitem{} Gregory, P.~C., Vavasour, J.~D., Scott, W.~K., \& Condon, J.~J. 1994, ApJS 90, 173
\bibitem{} Haigh, A. J. Robertson, J. G., \& Hunstead, R. W. 1997, PASA 14, 221
\bibitem{} Haigh, A. J. 2000, PhD thesis, University of Sydney
\bibitem{} Hambly, N. C., Davenhall, A. C., Irwin, M. J.,, \& MacGillivray, H. T. 2001, MNRAS, 326, 1315
\bibitem{} Hardcastle M.~J., Evans, D.~A. \& Croston, J.~H. 2007, MNRAS, 376, 1849
\bibitem{} Hummel, E. 1981, A\&A, 93, 93
\bibitem{} Jones, D. H., Saunders, W., Colless, M., et al.,
    2004, MNRAS, 355, 747 
\bibitem{} Jones, D. H., Saunders, W., Read, M., \& Colless M., 2005, PASA, 22, 277 
\bibitem{} Katgert, P., Mazure, A., den Hartog, R., Adami, C., Biviano, A., \&
    Perea, J., 1998, A\&AS, 129, 399
\bibitem{} Large, M.~I.,Cram, L.~E., \& Burgess, A.~M. 1991, MNRAS, 111, 72
\bibitem{} Ledlow, M. J., \& Owen, F. N. 1995a, AJ, 109, 853
\bibitem{} Ledlow, M. J., \& Owen, F. N. 1995b, AJ, 110, 1959
\bibitem{} Ledlow, M. J., \& Owen, F. N. 1996, AJ, 112, 9
\bibitem{} Ledlow, M. J., Voges, W., Owen, F. N., \& Burns, J. O. 2003, AJ, 126
\bibitem{} Lucey, J. R., Dickens, R. J., Mitchell, R. J., \& Dawe, J. A., 1983, MNRAS, 203, 5
\bibitem{} Mauch, T., Murphy, T., Buttery, H. J., et al.\ 2003, MNRAS, 342, 1117
\bibitem{} McHardy, I. M. 1979, MNRAS, 188, 495
\bibitem{} Mills, B. Y, \& Hoskins, D. G. 1977, AuJPh, 30, 509
\bibitem{} Nonino, M., Bertin, E., da Costa, L., Deul, E., et al.\ 1999, A\&AS 137, 51
\bibitem{} Owen, F. N., White, R. A. \& Burns, J. O. 1992, ApJS, 80, 501
\bibitem{} Owen, F. N., White, R. A. \& Ge, J.-P. 1993, ApJS, 87, 135
\bibitem{} Peacock, J.A. \& West, M., 1992, MNRAS, 259, 494
\bibitem{} Perley, R. A. 1979, AJ, 84, 1443
\bibitem{} Prandoni, I., Gregorini, L., Parma, P., de\,Ruiter, H. R., Vettolani, G., Wieringa, M. H.,
   \& Ekers, R. D., 2000, A\&AS, 146, 41   
\bibitem{} Prandoni, I., Gregorini, L., Parma, P., de\,Ruiter, H. R., Vettolani, G., Wieringa, M. H.,
   \& Ekers, R. D., 2001, A\&A, 365, 392   
\bibitem{} Robertson, J. G. \& Roach, G. J., 1990, MNRAS, 247, 387        
\bibitem{} Sault, R. J., Teuben, P. J., \& Wright, M. C. H. 1995, in ASP Conf.\ Ser.\ 77, 433, San~Francisco: ASP
\bibitem{} Schlegel, D. J., Finkbeiner, D. P., \& Davis, M. 1998, ApJ, 500, 525
\bibitem{} Slee, O. B. \& Siegman, B. C. 1983, Proc.\ ASA 5, 114
\bibitem{} Slee, O. B., Wilson, I. R. G. \& Siegman, B. C. 1983, AuJPh, 42, 633 
\bibitem{} Slee, O. B., Sadler, E. M., Reynolds, J. E., \& Ekers, R. D. 1994, MNRAS, 269, 928 
\bibitem{} Slee, O. B., Roy, A. L., \& Savage, A. 1994, AuJPh, 47, 145 
\bibitem{} Slee, O. B., Roy, A. L., \& Andernach, H. 1996, AuJPh, 49, 977 
\bibitem{} Slee, O. B., Roy, A. L., \& Andernach, H. 1998, AuJPh, 51, 971 
\bibitem{} Tyson, J. A., \& Jarvis, J. F. 1979, ApJ, 230, L153 
\bibitem{} Tsarevsky, G., et al.\ 2008, in preparation 
\bibitem{} Unewisse, A. M. 1993, PhD thesis, University of Sydney, Australia
\bibitem{} Verkhodanov, O. V., Trushkin, S. A., Andernach, H., \& Chernenkov, V. N. 1997,
          in ASP Conf.\ Ser.\ 125, 322, San~Francisco: ASP
\bibitem{} Wall, J.V. 1994, AuJPh, 47, 625
\bibitem{} West, R. M., \& Frandsen, S., 1981, A\&AS, 44, 329
\bibitem{} Wright, A.~E., Griffith, M.~R., Burke, B.~F., \&
        Ekers, R.~D. 1994, ApJS, 91, 111
\bibitem{} Zhao, J.-H., Burns, J. O. \& Owen, F. N. 1989, AJ, 98, 64
\end{thebibliography}


\setcounter{table}{1}
\begin{table*}
{\bf Table 2.} The list of radio sources detected in the seven cluster fields. \\
\begin{tabular}{lcccrrrrrrrcrr}
\hline
Source &  Radio~Centroid   & ang. & S   & $\Delta$S &   S & $\Delta$S &   S & $\Delta$S &  Spectral~~~ &  Sp. & \multicolumn{2}{c}{Deconvolved Structure} & \\
Name   &  RA, DEC (J2000)  & dist. & \multicolumn{2}{c}{0.843\,GHz} & \multicolumn{2}{c}{1.38\,GHz}  & \multicolumn{2}{c}{2.38\,GHz} &  index~~~~ &  cl. &  major~minor & P.A. & Note \\
       &  h~~~m~~~s~~~~~~~$^{\circ}~~~~'~~~~''$ & r/$R_A$ & mJy & mJy &  mJy & mJy &  mJy &  mJy & $\alpha$~~~~$\Delta\alpha$~~ &  &   $''$~~~~~~$''$  &  $^{\circ}$  &  \\
(1)         &      (2)~~~~~~(3)               & (4) &   (5) &  (6)  & (7)  &  (8)   & (9) & (10)  & (11)~~~(12) & (13) & (14)~~~~(15)&(16)&(17) \\
\hline 
A2746 ~1    & 00\,13\,09.69\,$-$65\,54\,05.0 & 0.90 &   3.0 & 1.5 &    4.04 & 0.33 &    2.77 &0.31 & $-$0.60 ~ .24~ &      &           &     &    \\
A2746 ~2a   & 00\,13\,24.18\,$-$65\,53\,45.3 & 0.87 &       &     &    1.94 & 0.37 & $<$1.5~ &     &                &      &           &     &    \\
A2746 ~2b   & 00\,13\,28.53\,$-$65\,53\,33.5 & 0.87 &       &     &    0.89 & 0.16 & $<$1.5~ &     &                &      &           &     &    \\
A2746 ~2a/b & 00\,13\,25.55\,$-$65\,53\,41.6 & 0.87 &   5.9 & 1.4 &    2.93 & 0.31 & $<$1.5~ &     & $-$1.42 ~ .53~ &      & 29.~ ~~~~~&  66 &    \\
A2746 ~3    & 00\,13\,32.83\,$-$65\,50\,36.3 & 1.06 &   6.9 & 1.4 &    3.63 & 0.34 & $<$1.5~ &     & $-$1.30 ~ .45~ &      &           &     &    \\
A2746 ~4    & 00\,13\,41.54\,$-$66\,16\,23.5 & 0.88 &   3.0 & 1.7 &    3.47 & 0.33 &    3.67 &0.42 & $+$0.12 ~ .26~ &      &           &     &    \\
A2746 ~5    & 00\,14\,16.20\,$-$66\,04\,21.2 & 0.03 &  12.1 & 1.6 &   10.0~ & 0.7~ &    8.2~ &0.5  & $-$0.38 ~ .13~ &      &  3.9 ~1.9 & 132 &    \\
A2746 ~6    & 00\,14\,17.91\,$-$66\,12\,52.5 & 0.59 &   2.8 & 1.4 &    1.56 & 0.19 &    1.09 &0.25 & $-$0.77 ~ .42~ &      &           &     &    \\
A2746 ~7    & 00\,14\,41.26\,$-$66\,19\,59.9 & 1.11 &  14.8 & 1.3 &   12.3~ & 0.8~ &    7.7~ &1.3  & $-$0.54 ~ .17~ &C$-$? &           &     &    \\
A2746 ~8    & 00\,15\,02.79\,$-$65\,50\,14.2 & 1.08 & 115.~ & 4.9 &   70.4~ & 4.2~ &   39.9~ &2.7  & $-$1.02 ~ .08~ &      &  5.2 ~2.6 &  41 &    \\
A2746 ~9    & 00\,15\,25.68\,$-$66\,19\,26.7 & 1.16 &  11.7 & 1.2 &    7.94 & 0.51 &    6.7~ &0.9  & $-$0.56 ~ .16~ & C+?  &  8.7 ~7.7 &  40 &    \\
A2746 10    & 00\,15\,34.31\,$-$66\,14\,20.3 & 0.88 &   6.8 & 1.3 &    4.61 & 0.32 &    2.92 &0.65 & $-$0.81 ~ .28~ &      &           &     &    \\
A2746 11    & 00\,15\,35.49\,$-$66\,08\,54.5 & 0.64 &   3.6 & 1.4 &    1.98 & 0.22 &    1.42 &0.39 & $-$0.81 ~ .43~ &      &           &     &    \\
A2746 12    & 00\,15\,45.26\,$-$66\,06\,35.0 & 0.64 &   6.8 & 1.2 &    2.85 & 0.29 &    2.26 &0.28 & $-$0.93 ~ .20~ & C+   &           &     &    \\
A2746 13    & 00\,15\,52.62\,$-$65\,51\,37.4 & 1.16 &   3.4 & 0.6 &    2.92 & 0.25 & $<$4.0~ &     & $-$0.31 ~ .40~ &      &           &     &    \\
A2746 14    & 00\,16\,22.87\,$-$66\,06\,39.2 & 0.91 &   3.3 & 1.1 &    1.89 & 0.24 &    2.06 &0.58 & $-$0.35 ~ .41~ & C+?  &           &     &    \\[1ex]
A2837 ~1    & 00\,46\,07.73\,$-$80\,13\,54.6 & 0.93 &  61.8 & 2.3 &   42.5~ & 2.6~ &   32.8  &2.45 & $-$0.63 ~ .07~ &      &  7.0 ~3.2 &  92 &  1 \\ 
A2837 ~2    & 00\,46\,35.33\,$-$80\,10\,53.0 & 0.90 &   8.2 & 1.0 &    5.17 & 0.44 &    3.64 &0.53 & $-$0.80 ~ .18~ &      &           &     &    \\
A2837 ~3    & 00\,48\,59.61\,$-$80\,09\,55.1 & 0.62 &   7.4 & 1.1 &    4.41 & 0.37 &    2.46 &0.45 & $-$1.06 ~ .23~ &      &           &     &    \\
A2837 ~4    & 00\,49\,34.52\,$-$80\,28\,00.7 & 0.79 & 157.~ & 5.~ &   86.2~ & 5.2~ &   46.2~ &3.1  & $-$1.13 ~ .06~ & C+?  &  5.8 ~2.2 &  18 &  2 \\ 
A2837 ~5    & 00\,49\,53.79\,$-$80\,00\,54.4 & 0.92 &  21.9 & 1.1 &   16.6~ & 1.1~ &    9.9~ &1.3  & $-$0.69 ~ .12~ &      &           &     &    \\
A2837 ~6    & 00\,50\,47.48\,$-$80\,15\,42.2 & 0.27 &  24.6 & 1.2 &   15.2~ & 0.9~ &    9.0~ &0.6  & $-$0.97 ~ .08~ &      & 12.~ ~3.1 &  39 &    \\
A2837 ~7    & 00\,50\,58.96\,$-$80\,32\,36.2 & 0.94 &  19.5 & 1.2 &   10.0~ & 0.8~ & $<$5.5~ &     & $-$1.35 ~ .20~ &      &           &     &    \\
A2837 ~8    & 00\,51\,01.98\,$-$80\,07\,54.1 & 0.50 &   6.0 & 1.4 &    2.40 & 0.16 & $<$1.2  &     & $-$1.86 ~ .49~ &      & 20.~ ~9.3 &  62 &    \\
A2837 ~9    & 00\,51\,15.82\,$-$80\,23\,26.2 & 0.46 &   9.2 & 1.1 &    5.47 & 0.47 &    3.49 &0.40 & $-$0.93 ~ .16~ &      &           &     &    \\
A2837 10    & 00\,51\,44.50\,$-$80\,16\,45.7 & 0.15 &   2.4 & 1.0 &    1.16 & 0.09 &    1.5  &0.2  &  +0.25: ~ .25~ & C+?  & 15.~ ~3.2 & 104 &  3 \\ 
A2837 11    & 00\,52\,02.70\,$-$80\,24\,40.9 & 0.49 &$<$3.8 &     &    2.25 & 0.28 & $<$1.5  &     &                &      & 24.~ ~9.6 &  96 &    \\
A2837 12    & 00\,52\,45.94\,$-$80\,15\,28.7 & 0.03 &   5.2 & 1.0 &    1.60 & 0.27 &    1.11 &0.17 & $-$1.41 ~ .23~ & C+   &           &     &    \\
A2837 13    & 00\,53\,30.04\,$-$80\,09\,47.8 & 0.35 &$<$3.8 &     &    1.21 & 0.16 &    1.98 &0.29 & $+$0.91 ~ .36~ &      &           &     &    \\
A2837 14    & 00\,53\,37.45\,$-$80\,00\,36.2 & 0.85 &   5.6 & 1.0 &    3.07 & 0.31 & $<$3.4~ &     & $-$1.22 ~ .42~ &      & 14.~ ~7.2 &  55 &    \\
A2837 15    & 00\,53\,53.04\,$-$79\,57\,29.5 & 1.02 &$<$3.8 &     &    3.43 & 0.58 & $<$5.5~ &     &                &      &           &     &    \\
A2837 16    & 00\,55\,02.02\,$-$80\,16\,47.1 & 0.32 &   8.3 & 0.9 &    4.98 & 0.34 &    3.41 &0.33 & $-$0.85 ~ .14~ &      &           &     &    \\
A2837 17    & 00\,57\,32.30\,$-$80\,26\,45.2 & 0.88 &   4.5 & 1.2 &    2.53 & 0.43 & $<$2.9~ &     & $-$1.2~ ~ .6~~~&      &           &     &    \\
A2837 18    & 00\,57\,49.26\,$-$80\,24\,18.8 & 0.83 &   9.3 & 1.1 &    5.23 & 0.40 & $<$3.6~ &     & $-$1.17 ~ .29~ &      &           &     &    \\
A2837 19    & 00\,58\,57.95\,$-$80\,22\,13.0 & 0.93 &$<$3.8 &     &    2.03 & 0.19 & $<$5.5~ &     &                &      &           &     &    \\[1ex]
A3126 ~1    & 03\,26\,07.45\,$-$55\,43\,34.8 & 0.91 &  41.9 & 1.7 &   18.2~ & 1.1~ &         &     & $-$1.69 ~ .15~ &      &           &     &    \\
A3126 ~2    & 03\,26\,22.90\,$-$55\,32\,33.9 & 0.93 & 224.~ & 7.  &  142.~~ & 9.~~ &   91.1~ &6.2  &~$-$0.91:~ .07~ &C$-$  & 11.~ ~3.8 & 135 &  4 \\ 
A3126 ~3    & 03\,26\,45.76\,$-$55\,47\,33.7 & 0.71 &   5.2 & 0.9 &    1.92 & 0.16 &         &     & $-$2.02 ~ .39~ &      &           &     &    \\
A3126 ~4    & 03\,26\,56.08\,$-$55\,41\,46.4 & 0.62 &$<$3.8 &     &    1.47 & 0.13 & $<$2.8~ &     &                &      &           &     &    \\
A3126 ~5    & 03\,27\,31.83\,$-$55\,24\,07.3 & 0.90 &   8.5 & 0.8 &    6.71 & 0.54 &         &     & $-$0.48 ~ .25~ &      &           &     &    \\
A3126 ~6    & 03\,27\,36.03\,$-$55\,33\,29.7 & 0.55 &  11.3 & 1.2 &    7.13 & 0.48 &    4.37 &0.41 & $-$0.92 ~ .14~ &      &           &     &    \\
A3126 ~7    & 03\,27\,50.50\,$-$55\,53\,33.5 & 0.55 &  14.8 & 1.1 &    6.93 & 0.68 &    3.63 &0.30 & $-$1.36 ~ .11~ &      &           &     &    \\
A3126 ~8    & 03\,27\,51.33\,$-$55\,56\,43.9 & 0.66 &  19.6 & 1.2 &   10.6~ & 0.7~ &    8.21 &0.63 & $-$0.87 ~ .09~ & C+?  &           &     &    \\
A3126 ~9    & 03\,27\,59.50\,$-$55\,25\,14.9 & 0.79 &   8.8 & 0.7 &    5.36 & 0.41 &         &     & $-$1.01 ~ .22~ &      & 13.~ ~9.8 &  38 &    \\
A3126 10a   & 03\,28\,08.08\,$-$55\,38\,20.2 & 0.26 &       &     &         &      &    3.82 &0.25 &                &      &           &     &    \\
A3126 10b   & 03\,28\,09.70\,$-$55\,38\,30.9 & 0.25 &       &     &         &      &    2.50 &0.19 &                &      &           &     &    \\
A3126 10a/b & 03\,28\,08.74\,$-$55\,38\,24.0 & 0.26 &  15.6 & 1.4 &    8.13 & 0.52 &    6.22 &0.47 & $-$0.85 ~ .11~ &      & 17.~ ~~~~~& 128 &    \\
A3126 11    & 03\,28\,17.80\,$-$55\,46\,26.9 & 0.20 &   5.1 & 0.7 &    1.13 & 0.15 &    1.26 &0.15 &~$-$1.26:~ .17~ & C+?  &           &     &  5 \\ 
A3126 12    & 03\,28\,25.41\,$-$55\,55\,52.3 & 0.57 &   5.2 & 0.8 &    2.36 & 0.21 &    1.38 &0.17 & $-$1.24 ~ .19~ &      &           &     &    \\
A3126 13    & 03\,28\,30.99\,$-$55\,42\,22.1 & 0.04 &       &     &   85.4~ & 4.3~ &   52.4~ &3.5~ & $-$0.90 ~ .16~ &      & 19.~ 11.~ & 115 &    \\
A3126 13+16 & 03\,28\,31.46\,$-$55\,42\,27.3 & 0.03 & 169.~ & 7.~ &   93.3~ & 5.9~ &   59.2~ &3.9~ & $-$1.03 ~ .08~ &      & 32.~ 14.~ & 125 &    \\
A3126 14    & 03\,28\,33.91\,$-$55\,49\,05.3 & 0.27 &   4.3 & 0.9 &    1.7~ & 0.14 &    1.52 &0.13 & $-$0.61 ~ .17~ & C+   &           &     &    \\
A3126 15    & 03\,28\,34.62\,$-$56\,04\,56.7 & 0.96 &  57.9 & 2.2 &   34.4~ & 2.1~ &         &     & $-$1.06 ~ .15~ &      & ~5.7 ~3.3 & 178 &    \\
A3126 16    & 03\,28\,35.83\,$-$55\,42\,44.4 & 0.01 &       &     &   27.0~ & 1.7~ &    6.67 &0.43 & $-$0.27 ~ .54~ &      &           &     &    \\
A3126 17    & 03\,28\,48.11\,$-$55\,47\,30.1 & 0.21 &   3.0 & 0.7 &    2.78 & 0.20 &    2.50 &0.17 & $-$0.19 ~ .16~ &      &           &     &    \\
A3126 18    & 03\,28\,52.66\,$-$55\,54\,45.2 & 0.52 &  10.6 & 0.9 &    7.91 & 0.52 &    6.09 &0.40 & $-$0.53 ~ .10~ &      &           &     &    \\
A3126 19    & 03\,29\,03.88\,$-$55\,59\,03.9 & 0.72 &   4.4 & 0.7 &    1.98 & 0.21 &         &     & $-$1.62 ~ .39~ &      &           &     &    \\
A3126 20    & 03\,29\,10.61\,$-$55\,38\,48.7 & 0.26 &$<$3.8 &     &    1.57 & 0.16 & $<$0.5~ &     &                &      &           &     &    \\
A3126 21    & 03\,29\,21.25\,$-$55\,48\,01.6 & 0.35 &$<$3.8 &     &    2.20 & 0.17 &    2.62 &0.19 & $+$0.32 ~ .20~ &      &           &     &    \\
A3126 22    & 03\,29\,40.65\,$-$55\,32\,34.1 & 0.58 & 208.~ & 7.~ &  150.~~ & 9.~~ &  113.~~ &7.~~ & $-$0.62 ~ .04~ &      & ~3.4 ~3.4 &     &  6 \\ 
A3126 23    & 03\,30\,00.79\,$-$55\,25\,38.9 & 0.90 &$<$3.8 &     &    2.28 & 0.24 &         &     &$>-$1.3~~~~~~~~ &      &           &     &    \\
\hline
\end{tabular}
\end{table*}
\begin{table*}
\newpage
{\bf Table 2.} (continued)
\begin{tabular}{lcccrrrrrrrcrr}
\hline
Source &  Radio~Centroid   & ang. & S   & $\Delta$S &   S & $\Delta$S &   S & $\Delta$S &  Spectral~~~ &  Sp. & \multicolumn{2}{c}{Deconvolved Structure} & \\
Name   &  RA, DEC (J2000)  & dist. & \multicolumn{2}{c}{0.843\,GHz} & \multicolumn{2}{c}{1.38\,GHz}  & \multicolumn{2}{c}{2.38\,GHz} &  index~~~~ &  cl. &  major~minor & P.A. & Note \\
       &  h~~~m~~~s~~~~~~~$^{\circ}~~~~'~~~~''$ & r/$R_A$ & mJy & mJy &  mJy & mJy &  mJy &  mJy & $\alpha$~~~~$\Delta\alpha$~~ &  &   $''$~~~~~~$''$  &  $^{\circ}$  &  \\
(1)         &      (2)~~~~~~(3)               & (4) &   (5) &  (6)  & (7)  &  (8)   & (9) & (10)  & (11)~~~(12) & (13) & (14)~~~~(15)&(16)&(17) \\
\hline
A3126 24    & 03\,30\,04.18\,$-$55\,29\,45.8 & 0.77 &   3.7 & 1.0 &    2.20 & 0.20 & $<$1.4~ &     & $-$1.05 ~ .58~ &      &           &     &    \\
A3126 25    & 03\,30\,10.29\,$-$55\,26\,04.1 & 0.92 &  47.5 & 2.0 &   27.9~ & 1.7~ &         &     & $-$1.08 ~ .15~ &      &  8.0 ~4.2 &  46 &    \\
A3126 26    & 03\,30\,26.02\,$-$55\,32\,43.6 & 0.79 &  22.4 & 1.2 &   11.5~ & 0.7~ &    8.39 &0.76 & $-$1.02 ~ .10~ &      &           &     &    \\
A3126 27    & 03\,30\,30.64\,$-$55\,55\,59.6 & 0.89 &  92.5 & 3.3 &   55.8~ & 3.4~ &         &     & $-$1.02 ~ .14~ &      & 22.~ ~2.6 & 126 &    \\
A3126 28    & 03\,31\,19.77\,$-$55\,48\,11.8 & 1.01 &   5.8 & 0.8 &    3.49 & 0.30 &         &     & $-$1.03 ~ .33~ &      & 13.~ ~1.8 &  55 &    \\
A3126 29    & 03\,31\,34.45\,$-$55\,33\,33.6 & 1.15 &  19.7 & 1.2 &   11.6~ & 0.9~ &         &     & $-$1.07 ~ .20~ &      &           &     &    \\[1ex]
A3216 ~1    & 04\,01\,49.18\,$-$65\,14\,19.7 & 1.04 &   3.4 & 1.7 &    2.04 & 0.31 &         &     & $-$1.04 ~ .06~ &      &           &     &    \\
A3216 ~2    & 04\,02\,21.46\,$-$65\,11\,40.9 & 0.79 &$<$12.~&     &    3.31 & 0.28 &    3.77 &0.40 & $+$0.24 ~ .25~ &      & 10.~ ~0.9 &  75 &    \\
A3216 ~3    & 04\,02\,29.43\,$-$65\,09\,44.3 & 0.76 &$<$16.~&     &    1.72 & 0.16 &    1.41 &0.25 & $-$0.37 ~ .37~ &      &           &     &    \\
A3216 ~4    & 04\,02\,49.81\,$-$65\,20\,14.8 & 0.80 &$<$11.~&     &    2.29 & 0.16 & $<$1.6~ &     &                &      &           &     &    \\
A3216 ~5    & 04\,03\,14.57\,$-$65\,06\,53.7 & 0.56 &$<$16.~&     &    1.48 & 0.13 &    1.12 &0.16 & $-$0.52 ~ .31~ &      &           &     &    \\
A3216 ~6a   & 04\,03\,15.21\,$-$65\,00\,29.7 & 0.94 &       &     &    5.89 & 0.38 &    2.91 &0.30 & $-$1.30 ~ .23~ &      & 11.~ ~2.0 &  21 &    \\
A3216 ~6b   & 04\,03\,13.50\,$-$65\,00\,53.4 & 0.92 &       &     &    3.18 & 0.24 &    0.98 &0.29 & $-$2.18 ~ .56~ &      & 13.~ ~7.3 &  45 &    \\
A3216 ~6a/b & 04\,03\,15.08\,$-$65\,00\,32.3 & 0.94 &  18.2 & 4.7 &    8.87 & 0.74 &    2.98 &0.35 & $-$1.89 ~ .22~ &      & 26.~ ~~~~~& 205 &    \\
A3216 ~7    & 04\,03\,34.45\,$-$65\,15\,49.1 & 0.34 &  79.2 & 7.5 &   35.4~ & 2.3~ &    8.37 &0.57 & $-$2.07 ~ .08~ &      & 34.~ ~6.~~&  60 &  7 \\ 
A3216 ~8    & 04\,04\,15.23\,$-$65\,21\,55.8 & 0.67 &$<$21.~&     &    4.89 & 0.47 & $<$1.6~ &     &                &      &           &     &    \\
A3216 ~9    & 04\,04\,22.62\,$-$65\,10\,12.8 & 0.20 &  50.0 & 8.2 &   33.7~ & 2.1~ &   21.3~ &1.4~ & $-$0.84 ~ .14~ &      &  6.0 ~4.2 & 156 &    \\
A3216 10    & 04\,04\,35.27\,$-$65\,03\,42.9 & 0.66 &$<$14. &     &    2.78 & 0.19 & $<$0.9~ &     &                &      & 12.~ ~4.1 &  82 &    \\
A3216 11    & 04\,05\,08.12\,$-$65\,14\,29.3 & 0.48 &  68.5 &11.~ &   26.5~ & 1.6~ &   19.0~ &1.3~ & $-$0.93 ~ .14~ & C+   &           &     &    \\
A3216 12    & 04\,05\,25.71\,$-$65\,23\,29.8 & 0.98 &  32.7 & 8.8 &   14.8~ & 0.9~ &    9.91 &0.85 & $-$0.86 ~ .17~ & C+   &           &     &    \\
A3216 13    & 04\,05\,37.01\,$-$65\,20\,38.4 & 0.89 &$<$19. &     &    5.20 & 0.35 &    4.20 &0.42 & $-$0.40 ~ .22~ &      &           &     &    \\
A3216 14    & 04\,05\,52.70\,$-$65\,01\,16.0 & 1.13 &$<$8.0 &     &    1.44 & 0.11 & $<$2.3~ &     &                &      &           &     &    \\
A3216 15    & 04\,05\,58.90\,$-$65\,06\,32.1 & 0.94 &$<$10.~&     &    8.26 & 0.51 &   20.1~ &1.5~ & $+$1.64 ~ .18~ &      &           &     &    \\
A3216 16    & 04\,06\,26.78\,$-$65\,12\,33.2 & 1.05 &$<$13.~&     &    4.04 & 0.27 &    3.20 &0.55 & $-$0.43 ~ .34~ &      &           &     &    \\[1ex]
A3230 ~1    & 04\,09\,30.78\,$-$63\,46\,20.4 & 0.94 &   8.9 & 2.4 &    4.63 & 0.50 & $<$1.4~ &     & $-$1.33 ~ .59~ &      & 12.6 ~6.9 & 143 &    \\
A3230 ~2    & 04\,09\,43.81\,$-$63\,48\,35.5 & 0.92 &  11.2 & 2.1 &    5.71 & 0.35 &    5.11 &0.36 & $-$0.44 ~ .14~ & C+   &           &     &    \\
A3230 ~3    & 04\,09\,57.36\,$-$63\,48\,55.3 & 0.84 &  11.2 & 2.0 &    6.21 & 0.42 &    4.48 &0.35 & $-$0.75 ~ .15~ &      &           &     &    \\
A3230 ~4    & 04\,10\,51.84\,$-$63\,44\,48.8 & 0.32 &   5.5 & 1.8 &    4.63 & 0.30 &    3.47 &0.24 & $-$0.52 ~ .16~ &      &           &     &    \\
A3230 ~5    & 04\,11\,13.05\,$-$63\,57\,15.8 & 1.12 &$<$3.8 &     &    2.43 & 0.20 & $<$1.7~ &     &                &      &           &     &    \\
A3230 ~6    & 04\,11\,18.86\,$-$63\,31\,48.7 & 0.72 &   4.2 & 1.3 &    1.57 & 0.11 & $<$0.5~ &     & $-$2.0  ~ .6~~ &      &           &     &    \\
A3230 ~7    & 04\,11\,26.64\,$-$63\,55\,56.0 & 1.03 &   9.0 & 1.4 &    4.64 & 0.39 &    2.81 &0.34 & $-$1.09 ~ .19~ &      &           &     &    \\
A3230 ~8    & 04\,11\,28.34\,$-$63\,41\,06.9 & 0.08 &  20.6 & 1.4 &    9.12 & 0.57 &    2.84 &0.19 & $-$1.92 ~ .09~ &      &  5.2 ~2.8 &  42 &    \\
A3230 ~9    & 04\,11\,31.21\,$-$63\,44\,09.6 & 0.19 &  31.5 & 1.8 &   22.9~ & 1.4~ &   13.2~ &0.8  & $-$0.84 ~ .08~ &      & 21.~ ~3.2 & 146 &    \\
A3230 10    & 04\,11\,35.93\,$-$63\,41\,00.9 & 0.14 &$<$3.8 &     & $<$1.0~ &      &    0.98 &0.12 &                &      & 17.~ ~3.2 & 113 &    \\
A3230 11    & 04\,11\,50.07\,$-$63\,47\,41.4 & 0.49 &$<$3.8 &     &    2.71 & 0.22 &    1.76 &0.13 & $-$0.80 ~ .20~ &      &           &     &    \\
A3230 12    & 04\,12\,01.52\,$-$63\,36\,48.6 & 0.49 &  28.6 & 1.5 &   21.2~ & 1.3~ &   13.7~ &0.8  & $-$0.71 ~ .08~ &      &  3.5 ~2.5 & 138 &    \\
A3230 13    & 04\,12\,03.35\,$-$63\,43\,43.8 & 0.37 &  93.1 & 3.4 &   57.4~ & 3.5~ &   28.5~ &1.7  & $-$1.13 ~ .07~ &      &  5.7 ~2.3 & 133 &    \\
A3230 14    & 04\,12\,06.99\,$-$63\,41\,58.0 & 0.37 &$<$3.8 &     &    4.07 & 0.26 &    3.05 &0.20 & $-$0.53 ~ .17~ &      &           &     &    \\
A3230 15    & 04\,12\,21.84\,$-$63\,34\,10.5 & 0.74 &  12.1 & 1.6 &    7.39 & 0.46 &    5.01 &0.33 & $-$0.80 ~ .12~ &      &           &     &    \\
A3230 16    & 04\,12\,25.92\,$-$63\,41\,15.3 & 0.53 &   7.8 & 1.2 &    3.05 & 0.21 &    0.99 &0.09 & $-$2.02 ~ .16~ &      &           &     &    \\
A3230 17    & 04\,13\,19.84\,$-$63\,36\,32.6 & 1.03 &   5.4 & 1.9 &    3.88 & 0.29 &    3.20 &0.23 & $-$0.39 ~ .18~ &      &           &     &    \\[1ex]
A3827 ~1    & 21\,59\,04.30\,$-$59\,57\,00.8 & 1.04 &  10.3 & 1.1 &    6.77 & 0.52 &         &     & $-$0.85 ~ .27~ &      &           &     &    \\
A3827 ~2    & 21\,59\,12.62\,$-$59\,46\,45.6 & 1.11 &  54.3 & 2.1 &   36.5~ & 2.3~ &         &     & $-$0.81 ~ .15~ &      &  8.4 ~4.5 & 117 &    \\
A3827 ~3    & 21\,59\,59.69\,$-$59\,39\,20.0 & 1.11 &  46.7 & 2.2 &   33.9~ & 2.4~ &         &     & $-$0.33 ~ .10~ & C+   &  8.7 ~4.6 &  39 &  8 \\ 
A3827 ~4    & 22\,00\,04.67\,$-$59\,49\,02.6 & 0.78 &   8.1 & 0.9 &    4.84 & 0.42 &    4.75 &0.60 & $-$0.54 ~ .16~ & C+   &           &     &    \\
A3827 ~5    & 22\,00\,34.62\,$-$59\,41\,01.6 & 0.92 &   8.8 & 1.4 &    2.44 & 0.23 &    6.75 &0.91 &  +0.00: ~ .20~ & C+?  &           &     &  9 \\ 
A3827 ~6    & 22\,00\,42.54\,$-$60\,07\,26.6 & 0.68 &  21.6 & 1.2 &   13.1~ & 0.8~ &    7.49 &0.84 & $-$1.02 ~ .11~ &      &  5.9 ~4.6 &  61 &    \\
A3827 ~7    & 22\,00\,42.62\,$-$60\,02\,56.0 & 0.53 &   5.4 & 0.9 &    2.31 & 0.20 &    2.02 &0.32 & $-$0.91 ~ .22~ & C+   &           &     &    \\
A3827 ~8    & 22\,01\,02.29\,$-$59\,43\,00.4 & 0.75 &$<$3.8 &     &    2.07 & 0.20 & $<$3.5~ &     &                &      & 20.~ 12.~~&   8 &    \\
A3827 ~9    & 22\,01\,03.27\,$-$59\,49\,54.0 & 0.47 &   6.4 & 1.0 &    3.26 & 0.29 &    2.52 &0.38 & $-$0.88 ~ .21~ &      & 13.~ ~3.7 &  55 &    \\
A3827 10    & 22\,01\,03.69\,$-$59\,51\,10.9 & 0.42 &$<$3.8 &     & $<$2.40 &      &    1.32 &0.25 &$<-$1.1~~~~~~   &      &           &     &    \\
A3827 11a   & 22\,01\,08.37\,$-$59\,40\,14.8 & 0.86 & 265.~ & 8.~ &  152.~~ & 9.~~ &   78.4~ &4.8  & $-$1.167~.052  &      &  9.2 ~3.9 &  36 & 10 \\ 
A3827 11b   & 22\,01\,11.59\,$-$59\,39\,23.2 & 0.90 & 204.~ & 8.~ &   91.4~ & 5.5~ &   33.9~ &2.2  & $-$1.72 ~ .07~ &      & 11.~ ~2.6 &  40 &    \\
A3827 11a/b & 22\,01\,09.44\,$-$59\,39\,57.5 & 0.87 & 469.~ &16.~ &  245.~~ &15.~~ &  111.~~ &7.   &~$-$1.12:~~.04: & Cpx  & 57.~ ~~~~~&  25 & 11 \\ 
A3827 12    & 22\,01\,17.07\,$-$59\,57\,55.1 & 0.24 &   2.6 & 0.9 &    0.88 & 0.13 &    0.70 &0.19 & $-$1.10 ~ .41~ &      &           &     &    \\
A3827 13    & 22\,01\,20.84\,$-$59\,34\,44.1 & 1.10 &$<$3.8 &     &    3.63 & 0.50 &         &     &$>-$0.12~~~~~~~ &      &           &     &    \\
A3827 14    & 22\,01\,35.46\,$-$60\,05\,17.2 & 0.42 &  23.5 & 1.2 &   10.5~ & 0.8~ &    7.00 &0.59 & $-$1.23 ~ .09~ & C+   & 20.~ ~6.~ &  59 &    \\
A3827 15    & 22\,01\,37.27\,$-$60\,08\,04.0 & 0.55 &$<$3.8 &     &    1.62 & 0.26 & $<$2.1~ &     &                &      &           &     &    \\
A3827 16    & 22\,01\,41.42\,$-$60\,02\,53.6 & 0.30 &$<$3.8 &     &    1.34 & 0.13 & $<$0.8~ &     &                &      &           &     &    \\
A3827 17    & 22\,01\,43.30\,$-$59\,56\,04.5 & 0.09 &$<$3.8 &     &    0.91 & 0.17 & $<$0.9~ &     &                &      &           &     &    \\
A3827 18    & 22\,01\,47.45\,$-$59\,51\,35.3 & 0.27 &$<$3.8 &     &    0.56 & 0.12 &    1.13 &0.22 & $+$1.30 ~ .54~ &      &           &     &    \\
\hline
\end{tabular}
\end{table*}
\begin{table*}
\newpage
{\bf Table 2.} (continued)
\begin{tabular}{lcccrrrrrrrcrr}
\hline
Source &  Radio~Centroid   & ang. & S   & $\Delta$S &   S & $\Delta$S &   S & $\Delta$S &  Spectral~~~ &  Sp. & \multicolumn{2}{c}{Deconvolved Structure} & \\
Name   &  RA, DEC (J2000)  & dist. & \multicolumn{2}{c}{0.843\,GHz} & \multicolumn{2}{c}{1.38\,GHz}  & \multicolumn{2}{c}{2.38\,GHz} &  index~~~~ &  cl. &  major~minor & P.A. & Note \\
       &  h~~~m~~~s~~~~~~~$^{\circ}~~~~'~~~~''$ & r/$R_A$ & mJy & mJy &  mJy & mJy &  mJy &  mJy & $\alpha$~~~~$\Delta\alpha$~~ &  &   $''$~~~~~~$''$  &  $^{\circ}$  &  \\
(1)         &      (2)~~~~~~(3)               & (4) &   (5) &  (6)  & (7)  &  (8)   & (9) & (10)  & (11)~~~(12) & (13) & (14)~~~~(15)&(16)&(17) \\
\hline
A3827 19    & 22\,01\,49.68\,$-$59\,58\,35.9 & 0.09 &   7.9 & 0.9 &    2.29 & 0.18 & $<$0.8~ &     & $-$2.51 ~ .28~ &      &           &     &    \\
A3827 20    & 22\,01\,52.17\,$-$60\,13\,42.1 & 0.81 &   5.4 & 0.9 &    1.87 & 0.32 &         &     & $-$2.15 ~ .48~ &      &           &     &    \\
A3827 21    & 22\,01\,57.74\,$-$59\,42\,52.6 & 0.68 &$<$3.8 &     &    2.34 & 0.32 & $<$1.8~ &     &                &      & 39.~ ~6.9 &  15 &    \\
A3827 22    & 22\,02\,00.72\,$-$59\,52\,57.0 & 0.20 &   6.6 & 1.1 &    5.15 & 0.42 &    3.70 &0.33 & $-$0.58 ~ .16~ &      & 10.0 ~5.2 &  54 &    \\
A3827 23    & 22\,02\,02.50\,$-$59\,55\,35.7 & 0.08 &   6.9 & 1.1 &    2.51 & 0.24 & $<$1.0~ &     & $-$2.05 ~ .38~ &      &           &     &    \\
A3827 24a   & 22\,02\,10.83\,$-$60\,09\,12.4 & 0.60 &       &     &    5.12 & 0.51 & $<$2.8~ &     &$<+$0.43~~~~~~  &      &           &     &    \\
A3827 24b   & 22\,02\,14.56\,$-$60\,08\,42.5 & 0.58 &       &     &    3.48 & 0.45 & $<$2.8~ &     &$<+$1.2~~~~~~~  &      &           &     &    \\
A3827 24a/b & 22\,02\,12.35\,$-$60\,09\,00.3 & 0.59 &  21.1 & 2.8 &    8.45 & 0.82 & $<$2.8~ &     & $-$1.86 ~ .33~ &      & 41.~ ~~~~~&  43 &    \\
A3827 25    & 22\,02\,38.73\,$-$59\,58\,57.3 & 0.28 &$<$3.8 &     & $<$1.88 &      &    1.50 &0.16 &                &      &           &     &    \\
A3827 26    & 22\,02\,55.71\,$-$60\,17\,02.6 & 1.04 & 279.  &10.4 &  157.~~ & 9.6~ &         &     & $-$0.99 ~ .08~ &      & 22.~ ~3.4 & 139 & 12 \\ 
A3827 27    & 22\,03\,18.50\,$-$60\,04\,22.4 & 0.62 &   7.3 & 0.9 &    4.05 & 0.39 &    2.77 &0.63 & $-$1.02 ~ .23~ &      & 13.~ ~2.9 &  88 &    \\
A3827 28    & 22\,03\,38.12\,$-$59\,43\,58.7 & 0.89 &  13.2 & 1.4 &    5.92 & 0.54 &         &     & $-$1.63 ~ .28~ &      &           &     &    \\
A3827 29    & 22\,03\,38.52\,$-$59\,41\,37.7 & 0.97 &  10.5 & 1.0 &    4.61 & 0.56 &         &     & $-$1.67 ~ .31~ &      &           &     &    \\
A3827 30    & 22\,04\,34.32\,$-$59\,56\,26.4 & 0.96 &   6.3 & 0.6 &    3.93 & 0.64 &         &     & $-$0.96 ~ .38~ &      &           &     &    \\[1ex]
A3836 ~1    & 22\,07\,18.96\,$-$51\,47\,34.4 & 1.03 &  11.0 & 1.1 &    6.44 & 0.48 &         &     & $-$1.09 ~ .25~ &      &           &     &    \\
A3836 ~2    & 22\,07\,51.85\,$-$51\,57\,12.8 & 0.88 &   6.1 & 1.2 &    2.80 & 0.27 &         &     & $-$1.58 ~ .45~ &      &           &     &    \\
A3836 ~3    & 22\,08\,02.31\,$-$51\,47\,31.6 & 0.67 &$<$3.8 &     &    1.52 & 0.22 & $<$1.4~ &     &                &      &           &     &    \\
A3836 ~4    & 22\,08\,33.58\,$-$51\,43\,50.6 & 0.49 &  11.4 & 1.2 &    6.83 & 0.49 &    4.20 &0.42 & $-$0.96 ~ .14~ &      &           &     &    \\
A3836 ~5    & 22\,08\,39.71\,$-$52\,02\,08.5 & 0.79 &  19.5 & 1.4 &    9.31 & 0.62 &    5.11 &0.72 & $-$1.36 ~ .14~ &      &           &     &    \\
A3836 ~6    & 22\,08\,51.91\,$-$51\,37\,01.4 & 0.69 &$<$3.8 &     &    1.66 & 0.14 & $<$1.8~ &     &                &      &           &     &    \\
A3836 ~7    & 22\,08\,58.47\,$-$51\,45\,09.1 & 0.29 &  10.5 & 1.4 &    8.97 & 0.57 &    5.09 &0.39 & $-$0.82 ~ .14~ &      &           &     &    \\
A3836 ~8    & 22\,09\,12.57\,$-$51\,30\,57.0 & 0.96 &  16.5 & 1.4 &    7.22 & 0.58 &         &     & $-$1.68 ~ .24~ &      &           &     &    \\
A3836 ~9a   & 22\,09\,19.01\,$-$51\,42\,55.7 & 0.32 &       &     &    4.61 & 0.38 &    2.68 &0.22 & $-$1.00 ~ .21~ &      &           &     &    \\
A3836 ~9b   & 22\,09\,17.07\,$-$51\,43\,24.0 & 0.30 &       &     &    2.51 & 0.32 &    1.76 &0.18 & $-$0.66 ~ .30~ &      &           &     &    \\
A3836 ~9a/b & 22\,09\,18.28\,$-$51\,43\,06.3 & 0.31 &  15.3 & 1.3 &    6.41 & 0.80 &    4.23 &0.33 & $-$1.23 ~ .11~ &      & 34.~ ~~~~~& 213 &    \\
A3836 10    & 22\,09\,18.66\,$-$52\,10\,14.8 & 1.14 &  11.7 & 1.2 &    7.04 & 0.62 &         &     & $-$1.03 ~ .27~ &      &           &     &    \\
A3836 11    & 22\,09\,19.07\,$-$51\,50\,54.9 & 0.11 &  15.0 & 1.4 &    8.15 & 0.52 &    4.21 &0.38 & $-$1.23 ~ .12~ &      &           &     &    \\
A3836 12    & 22\,09\,22.96\,$-$51\,56\,05.7 & 0.38 &   2.5 & 0.7 &    0.84 & 0.13 & $<$1.0~ &     & $-$2.2~ ~ .7~~~&      &           &     &    \\
A3836 13    & 22\,09\,26.52\,$-$51\,49\,12.4 & 0.03 &$<$3.8 &     &    1.07 & 0.16 & $<$0.8~ &     &                &      &           &     &    \\
A3836 14    & 22\,09\,32.15\,$-$51\,37\,30.4 & 0.61 &$<$3.8 &     &    0.95 & 0.17 & $<$1.8~ &     &                &      &           &     &    \\
A3836 15    & 22\,09\,52.28\,$-$52\,04\,40.7 & 0.88 &   4.5 & 0.9 &    1.31 & 0.16 & $<$3.2~ &     & $-$2.50 ~ .47~ &      &           &     &    \\
A3836 16    & 22\,09\,53.81\,$-$51\,48\,21.3 & 0.25 &$<$3.8 &     &    1.21 & 0.13 &    0.83 &0.17 & $-$0.70 ~ .43~ &      &           &     &    \\
A3836 17    & 22\,09\,57.09\,$-$51\,42\,30.5 & 0.44 &   4.4 & 1.4 &    2.02 & 0.23 & $<$1.2~ &     & $-$1.58 ~ .69~ &      & 34.~ ~5.1 & 140 &    \\
A3836 18    & 22\,10\,03.40\,$-$51\,53\,14.3 & 0.40 &   2.7 & 0.8 &    1.09 & 0.12 &    1.45 &0.26 & $-$0.18 ~ .30~ & C+   &           &     & 13 \\ 
A3836 19a   & 22\,10\,05.36\,$-$51\,40\,09.0 & 0.58 &       &     &   26.6~ & 1.6~ &   14.7~ &1.0  & $-$1.10 ~ .17~ &      &  6.2 ~4.3 & 147 &    \\
A3836 19b   & 22\,10\,02.26\,$-$51\,39\,36.0 & 0.59 &       &     &   13.5~ & 0.9~ &    5.57 &0.64 & $-$1.64 ~ .25~ &      & 17.~ ~3.1 & 158 &    \\
A3836 19a/b & 22\,10\,04.43\,$-$51\,39\,59.0 & 0.59 &  59.2 & 5.6 &   39.7~ & 2.4~ &   19.5~ &1.3  & $-$1.12 ~ .11~ &      & 44.~ ~~~~~& 319 &    \\
A3836 20    & 22\,10\,07.88\,$-$51\,55\,07.0 & 0.50 &$<$3.8 &     &    1.46 & 0.18 &    1.00 &0.17 & $-$0.70 ~ .39~ &      &           &     &    \\
A3836 21    & 22\,10\,12.25\,$-$52\,00\,06.8 & 0.72 &   4.7 & 1.2 &    3.62 & 0.35 &    2.40 &0.44 & $-$0.68 ~ .29~ &      &           &     &    \\
A3836 22    & 22\,10\,59.46\,$-$51\,42\,27.6 & 0.87 &   4.6 & 1.0 &    2.36 & 0.26 & $<$2.8~ &     & $-$1.35 ~ .49~ &      & 11.~ ~4.0 & 123 &    \\
A3836 23    & 22\,11\,03.67\,$-$51\,35\,39.6 & 1.09 &   4.2 & 1.1 &    2.18 & 0.25 &         &     & $-$1.33 ~ .58~ &      &           &     &    \\
A3836 24    & 22\,11\,15.03\,$-$51\,58\,00.5 & 1.04 &  80.1 & 2.8 &   37.5~ & 2.4~ &   18.8~ &2.8  & $-$1.47 ~ .11~ &      &  7.9 ~2.0 & 121 &    \\
\hline
\end{tabular}
\begin{flushleft}
Description of columns:
(1) Source name, composed of the Abell cluster name, followed 
   by a sequence number. For multiple component sources we list both individual components 
   and integrated parameters. 
(2,3) Right ascension and declination (equinox J2000.0) for the radio centroid.
(4) Projected angular distance of the source centroid in units of Abell radii (cf.\ Table~1).
(5,6) Total flux density at 0.843\,GHz and its error.
(7,8) Total flux density at 1.38\,GHz and its error.
(9,10) Total flux density at 2.38\,GHz and its error. 
(11,12) Spectral index ($S\propto\nu^{\alpha}$) and its error. These are based on either
  2-frequency pairs (0.843--1.38, 0.843--2.38 or 1.38--2.38\,GHz), or linear
     fits based on up to five frequency bands, the others being 0.408~(MRC),
      4.8~(PMN/PMNAT) and 8.64~GHz (PMNAT).  A colon after the spectral index indicates
     an uncertain value.
(13) Radio spectral shape, if available and different from a straight power law (see Sect.~4.1).
(14,15) Deconvolved major and minor axis of an elliptical fit to the source, in arcsec,
from either the 1.38 or the 2.38-GHz maps, see Sect.\ 3.2.
(16) Position angle of source major axis (N through E), in degrees.
(17) Notes on other names or spectral shape:  
 1: PMN\,J0045$-$8014;
 2: PMN\,J0049$-$8027;
 3: S(1.38MHz) too low? variable?;
 4: PMN and PMNM\,J0326$-$5532; PMNAT;
 5: S(1.38MHz) too low? variable?;
 6: PMN\,J0329$-$5532; PMNAT;
 7: MC4\,0403$-$653;
 8: PMN\,J2159$-$5939;
 9: S(1.38MHz) too low? variable?;
10: MRC\,2157$-$599; PMN\,J2201$-$5940; PMNAT;
11: MRC\,2157$-$599; PMN\,J2201$-$5940;
12: PMN\,J2202$-$6017;
13: S(1.38MHz) too low? variable?  References to acronyms are:
MRC\, = Large et al. (1991); MC4  = Clarke et al. (1976);
PMN\, = Wright et al. (1994); PMNM = Gregory et al.\ (1994);
PMNAT = Wright et al.\ (1997, ftp.atnf.csiro.au/pub/data/pmn/CA). Note that
we do NOT quote the SUMSS catalogue names here.
\end{flushleft}

\end{table*}


\setcounter{table}{2}
\begin{table*}
\caption{Optical identifications, radio powers, absolute magnitudes and notes.}
\begin{tabular}{lcccrrcccl}
\hline
Source   &  \multicolumn{6}{c}{Optical Objects near Radio Source}   \\
Name     &  Morph.  &  UKST  & UKST & Dist. & P.A.  &  $z_{\rm LG}$ & log\,P$_{1.38}$ & $M_R$ & Note \\
         &  Class   & $B_J$  &  R  & $''$  & $^{\circ}$~~ &  &  W\,Hz$^{-1}$  &   mag   & \\
       (1)   &  (2)  & (3)   &  (4) & (5) & (6) &    (7)    & (8)     &   (9)      & (10)  \\
\hline
A2746 ~~2a/b &  G   & 22.04 & 19.61 &  7 & 240 &            &          &             & *  \\
A2746 ~~5    &  cD  & 17.60 & 15.89 &  0 &     &            & 23.69~   & $-$23.71    &    \\
A2746 ~~8    &  St  & 21.62 &       &  1 &  68 &            &          &             & *  \\
A2746 11     &  G   & 21.07 & 18.66 &  2 & 201 &            & 22.96~   & $-$20.94    &    \\
A2746 13     &  E   & 19.93 & 17.95 &  1 & 139 &            & 23.16~   & $-$21.65    &    \\
A2837 ~~1    &  E   & 20.80 & 18.92 &  1 &  46 &            &          &             & *  \\
A2837 ~~3    &  St  & 21.83 & 19.69 &  5 &  66 &            &          &             &    \\
A2837 ~~6    &  G   & 19.97 & 18.22 &  3 & 286 &            & 23.55~   & $-$20.64    & *  \\
A2837 ~~8    &  B   &       &       &    &     &            &          &             & *  \\
A2837 ~~9    &  G   & 21.49 & 19.30 &  2 &  98 &            &          &             & *  \\
A2837 10     &  G   & 21.67 & 19.65 &  1 & 326 &            &          &             & *  \\
A2837 11     &  G   & 20.06 & 19.47 &  8 &  68 &            &          &             & *  \\
A2837 12     &  E   & 18.61 & 16.91 &  1 & 287 &            & 22.55~   & $-$21.95    &    \\
A2837 14     &  G   & 21.24 & 19.08 &  4 &  56 &            &          &             & *  \\
A2837 16     &  G   & 21.56 & 19.87 &  0 &     &            &          &             &    \\
A2837 17     &  G   & 21.56 & 19.55 &  3 & 270 &            &          &             &    \\
A2837 19     &  St  & 18.94 & 18.74 &  3 & 119 &            &          &             &    \\
A3126 ~~2    &  B   &       &       &    &     &            &          &             & *  \\
A3126 ~~8    &  St  & 21.83 & 19.56 &  1 &  94 &            &          &             &    \\
A3126 ~~9    &  D   & 15.70 & 14.19 &  1 &  58 &            & 22.85~   & $-$23.84    & *  \\ 
A3126 10a/b  &  B   &       &       &    &     &            &          &             & *  \\
A3126 11     &  2xE & 17.06 & 16.17 &  5 &  94 &            & 22.17~   & $-$21.86    &    \\
A3126 12     &  G   & 21.05 & 19.72 &  3 &  88 &            &          &             &    \\
A3126 13     &  E   & 16.86 & 15.45 &  6 & 345 & .0849(1,2) & 23.99~   & $-$22.58    & *  \\
A3126 16     &  E   & 16.49 & 15.11 &  1 & 109 & .0833(1,2) & 23.50~   & $-$22.92    & *  \\
A3126 17     &  St  & 19.29 & 18.78 &  1 & 144 &            &          &             &    \\
A3126 19     &  St  & 21.04 & 20.03 &  3 & 318 &            &          &             &    \\
A3126 20     &  E   & 18.63 & 17.27 &  8 & 190 & .0785 (2)  & 22.32~   & $-$20.76    &    \\
A3126 22     &  G   & 19.19 & 17.87 &  6 & 275 &            &          &             &    \\
A3126 25     &  St  & 20.54 & 18.50 &  6 & 341 &            &          &             & *  \\
A3126 27     &  G   & 21.26 & 19.47 &  1 & 324 &            &          &             & *  \\
A3126 28     &  G   & 20.74 & 18.74 &  5 & 200 &            &          &             & *  \\
A3216 ~~2    &  E   & 18.84 & 17.57 &  1 & 279 &            & 23.24~   & $-$22.09    & *  \\
A3216 ~~4    &  E   & 18.69 & 17.29 &  1 &  13 &            & 23.07~   & $-$22.37    &    \\
A3216 ~~6a/b &  G   & 20.02 & 17.84 &  1 & 290 &            & 23.54~   & $-$21.82    & *  \\
A3216 ~~7    &  G   & 21.02 & 19.26 & 16 &  64 &            &          &             & *  \\
A3216 ~~9    &  G   & 18.05 & 16.44 &  2 & 237 & .1581 (3)  & 24.18~   & $-$23.22    & *  \\
A3216 10     &  G   & 19.25 & 17.52 &  9 &  91 &            & 23.09~   & $-$22.14    & *  \\
A3216 14     &  B   &       &       &    &     &            &          &             & *  \\
A3216 15     &  St  & 19.56 & 18.28 &  1 & 287 &            &          &             &    \\
A3230 ~~1    &  E/S0& 18.60 & 17.11 & 21 & 126 &            & 23.31~   & $-$22.57    & *  \\
A3230 ~~2    &  G   & 19.87 & 18.37 &  4 & 287 &            & 23.46~   & $-$21.31    &    \\
A3230 ~~4    &  E   & 18.32 & 16.56 &  3 & 245 &            & 23.36~   & $-$23.12    & *  \\
A3230 ~~5    &  E   & 20.38 & 18.33 &  3 & 277 &            & 23.05~   & $-$21.35    &    \\
A3230 ~~7    &  E   & 20.39 & 19.00 &  3 & 234 &            & 23.33~   & $-$20.68    &    \\
A3230 ~~8    &  cD  & 17.51 & 15.68 &  1 & 222 &            & 23.57~   & $-$24.00    &    \\
A3230 ~~9    &  St  & 22.14 & 20.32 &  6 & 326 &            &          &             & *  \\
A3230 11     &  E   & 19.12 & 17.08 &  1 & 170 &            & 23.11~   & $-$22.60    & *  \\
A3230 14     &  G   & 21.20 & 18.66 &  1 & 267 &            & 23.31~   & $-$21.02    &    \\
A3827 ~~2    &  B   &       &       &    &     &            &          &             & *  \\
A3827 ~~3    &  St  & 19.73 & 18.10 &  3 &  77 &            &          &             & *  \\
A3827 ~~4    &  E   & 17.61 & 15.99 &  4 & 336 &            & 22.95~   & $-$22.44    & *  \\
A3827 ~~5    &  E   & 19.86 & 17.87 &  3 &  17 &            & 22.67~   & $-$20.56    &    \\
A3827 ~~8    &  E   & 18.09 & 16.80 & 11 & 343 &            & 22.65~   & $-$22.36    & *  \\
A3827 10     &  E   & 17.60 & 16.07 &  1 &  82 & .0967 (4)  & 22.41~   & $-$22.36    &    \\
A3827 11a/b  &  B   &       &       &    &     &            &          &             & *  \\
A3827 12     &  2xE & 18.70 & 16.53 &  3 & 290 & .0931 (4)  & 22.19~   & $-$21.90    &    \\
A3827 14     &  B   &       &       &    &     &            &          &             & *  \\
A3827 18     &  G   & 22.12 & 20.78 &  3 &  41 &            &          &             &    \\
A3827 20     &  E/S0& 18.41 & 16.95 &  3 & 348 &            & 22.47~   & $-$21.48    &    \\
A3827 21     &  G   & 21.35 & 18.61 &  5 &  33 &            &          &             & *  \\
\hline
\end{tabular}
\end{table*}
\begin{table*}
\newpage
\begin{tabular}{lcrrrrcccl}
{\bf Table 3.} (continued) \\
\hline
Source   &  \multicolumn{6}{c}{Optical Objects near Radio Source}   \\
Name     &  Morph.  &  UKST  & UKST & Dist. & P.A.  &  $z_{\rm LG}$ & log\,P$_{1.38}$ & $M_R$ & Note \\
         &  Class   & $B_J$  &  R  & $''$  & $^{\circ}$~~ &  &  W\,Hz$^{-1}$  &   mag   & \\
       (1)   &  (2)  & (3)   &  (4) & (5) & (6) &    (7)    & (8)     &   (9)      & (10)  \\
\hline
A3827 22     &  G   & 21.21 & 18.60 &  1 & 308 &            &          &             &    \\
A3827 23     &  St  & 22.13 & 19.85 &  2 &  51 &            &          &             &    \\
A3827 24a/b  &  Sp  & 19.06 & 17.97 & 30 &  46 &            &          &             & *  \\
A3827 25     &  E   & 18.07 & 16.59 &  1 &  22 & .0988 (4)  & 22.46~   & $-$21.84    &    \\
A3827 26     &  B   &       &       &    &     &            &          &             & *  \\
A3827 30     &  St  & 22.29 & 20.47 &  4 &  67 &            &          &             &    \\
A3836 ~~1    &  St  & 21.20 & 19.46 &  5 & 304 &            &          &             &    \\
A3836 ~~4    &  Sp  & 17.40 & 16.39 &  0 &     & .0647(5,6) & 22.73\,n & $-$20.99\,n &    \\
A3836 ~~8    &  St  & 19.93 & 18.40 &  4 & 168 &            &          &             &    \\
A3836 ~~9a/b &  B   &       &       &    &     &            &          &             & *  \\
A3836 13     &  E   &   ?   & 16.92 &  2 & 249 &            & 22.39~   & $-$21.73    &    \\
A3836 16     &  Sp  & 16.90 & 15.84 &  2 &  59 & .0365 (6)  & 21.49\,n & $-$20.21\,n &    \\
A3836 17     &  B   &       &       &    &     &            &          &             & *  \\
A3836 18     &  G   & 20.46 & 18.94 &  2 & 310 &            &          &             &    \\
A3836 19a/b  &  G   & 21.66 & 19.52 & 15 & 311 &            &          &             & *  \\
A3836 22     &  D   &   ?   & 15.58 &  3 &  17 &            & 22.70~   & $-$23.07    &    \\ 
\hline                                                                               
\end{tabular}                                                                        

\begin{flushleft} 
Description of columns:
(1) Source name, composed of Abell cluster name, followed by a sequence number. 
Multiple-component sources are listed only with one entry for the integrated source.
(2) morphological class of the optical identification (ID): St = stellar,
G = galaxy of unknown type, E = elliptical galaxy, S0 = S0 galaxy, Sp = spiral galaxy,
D= D galaxy; cD = cD galaxy; 2xE = two adjacent elliptical galaxies; B = blank field;
(3) $B_J$~magnitude from SuperCOSMOS database; a question mark indicates a
bright object but an unreliable value of $B_J$;
(4) $R$~magnitude from SuperCOSMOS database; The UK Schmidt red magnitudes are listed under R2 in
SuperCOSMOS. R1 refers to ESO.
(5) angular distance of the optical ID from the radio source position, in arcsec;
(6) position angle of the radius vector from the radio source to the optical ID, from
N through E;
(7) redshift of the optical ID in the galactocentric frame, i.e.\ 
z$_{LG}$= c\,z$_{hel}$+\,300\,km\,s$^{-1}$~sin$\ell$\,cos\,b, with numbers in brackets referrring
to the following references:
1. Lucey et al.\ (1983); 2. Colless \& Hewett (1987); 3. Jones et al.\ (2004);
4. Katgert et al.\ (1998); 5. de\,Grijp et al.\ (1992); 6. Ebeling (1997); 
(8) log$_{10}$ of the 1.38-GHz radio power in W\,Hz$^{-1}$, assuming the 
redshift of the cluster as indicated in Table~1; for IDs not considered cluster 
members the value is appended by a letter ``n''.
(9) absolute $R$~magnitude for the optical ID, assuming the redshift of
the cluster as indicated in Table~1.  Note that galaxy identifications are
considered as cluster members only if their absolute R-band magnitudes,
when obtained by applying the cluster redshift (in Table~1) to the apparent
red magnitude, is brighter than $-$20.5. For IDs not considered cluster members 
the value is appended by a letter ``n''.  
(10) Notes to individual sources, see Section~3.4. \\

\end{flushleft}

\vspace*{5cm}

\centerline{\Huge\bf Online Material}

\vspace*{2cm}

Cleaned 1.38-GHz maps of A2746, A2837, A3126, A3216, A3230 and A3827
(uncorrected for primary beam attenuation) centred on the X-ray centroid
marked ``X'', and covering an area of between 36$'\times$36$'$ and 48$'\times$48$'$.
The dashed circles mark the Abell radius (2.0\,Mpc for the adopted cosmology,
cf.\ also Table 1), and the radio sources are numbered according to their
entries in Tables~2 and 3.  The size and orientation of the restoring beam
is shown in the lower left corner and listed numerically underneath each
figure, as is the level of the lowest and highest contours. 
The rms noise over clear areas near the centre is listed in the
penultimate column of Table~1. The primary beam width (FWHM) at 1.38~GHz
is 32$'$.

\end{table*}



\end{document}